\documentclass[a4paper, 11pt, oneside]{Thesis}  
\graphicspath{{./Figures/}}  

\usepackage[square, numbers, comma, sort&compress]{natbib}  
\usepackage{hypernat}
\usepackage{verbatim}  
\usepackage{vector}  
\usepackage{dsfont}
\hypersetup{urlcolor=MidnightBlue, colorlinks=true}  
\usepackage{enumerate}
\usepackage{booktabs}
\usepackage{multirow}
\usepackage{multicol}
\usepackage{rotating} 
\usepackage{graphics}

\begin{document}
\frontmatter	  

\title  {Transaction Costs in Execution Trading}
\authors  {\texorpdfstring
            {\href{https://www.linkedin.com/in/david-marcos-74192453/}{David Marcos}}
            {David Marcos}
            }         
\addresses  {\groupname\\\deptname\\\univname}  
\date {December 2019}
\subject    {}

\maketitle

\newpage{\
\thispagestyle{empty}}
\newpage{\
\thispagestyle{empty}}

\setstretch{1.3}  

\fancyhead{}  
\rhead{\thepage}  
\lhead{}  

\pagestyle{fancy}  

\pagestyle{empty}  

\null\vfill

\begin{flushright}

\textit{We shall not cease from exploration,\\
and the end of all our exploring\\ 
will be to arrive where we started\\ 
and know the place for the first time.}

T. S. Elliot
\end{flushright}

\vfill\vfill\vfill\vfill\vfill\vfill\null
\clearpage  


%
%

~

~

~

~

\begin{flushleft}

{\it ``This too to remember. If a man writes clearly enough any one can see if he fakes. If he mystifies to avoid a straight statement, which is very different from breaking so-called rules of syntax or grammar to make an effect which can be obtained in no other way, the writer takes a longer time to be known as a fake and other writers who are afflicted by the same necessity will praise him in their own defense. True mysticism should not be confused with incompetence in writing which seeks to mystify where there is no mystery but is really only the necessity to fake to cover lack of knowledge or the inability to state clearly. Mysticism implies a mystery and there are many mysteries; but incompetence is not one of them; nor is overwritten journalism made literature by the injection of a false epic quality. Remember this too: all bad writers are in love with the epic.''}

\end{flushleft}

\begin{flushleft}
Ernest Hemingway.
\end{flushleft}




\clearpage  

%
%
%
%
%
%
%


\acknowledgements{
\addtocontents{toc}{\vspace{0em}}  

\begin{flushright}

\textit{``For a successful technology,\\ reality must take precedence\\ over public relations,\\ for Nature cannot be fooled.''}

Richard P. Feynman

\end{flushright}

~

Writing acknowledgements under the regulatory constraint of not making it possible to identify the writer is certainly a challenging task. However, I realise that the people on this list are extremely unique. So I hope this makes my vagueness not challenging for them when trying to identify themselves.

~

First and foremost, I would like to thank my Thesis advisor. Not only for providing me guidance, but also for allowing me the intellectual freedom that was necessary to complete this work. Finding someone that combines a considerable degree of both wisdom and modesty is, in my opinion, a lucky rare event. My advisor is one of them. 

~

The peculiar members of the trading team deserve a special mention, starting with P. Hegetschweiler, M. Both, A. Papale, and C. Laske; and followed by L. Challenger and M. David, among others, who have massively contributed to my knowledge of how trading is done in the real world (and about recommended wine, restaurants, and other amenities). 

~

I am still indebted to my advisor in Physics, Ram\'on A. --from whom I learned one of the best abilities that a man can have: critical thinking; and as a corollary the science of academic writing. To this day, his human and professional abilities remain an inspiration for me.

~

At Oxford I left many unforgettable memories and people. Thanks to the long list of classmates with whom I shared fun and desperation. Thanks to the Maths Department, both Professors and Secretaries have made the Programme extremely well organised and easy to follow (despite of the sometimes long pricing formulas). Thanks to that Professor that helped me land my first job in Finance. Thanks to St. Anne's College, for being both liberal in spirit and deep in nature. Thanks to my friends of adventure at Oxford, mainly to Mati Kaim and Jasper Beerepoot, who both showed me that crisis and opportunity can be synonymous words. 

~

This Thesis would not exist without my closest friends and family. These are the people that have been next to me day after day, on sunny or rainy days. Thank you, Alejandro G. Alfonso, for being as close as a brother to me. Sara S. Bailador, our childhood really was the beginning of a beautiful friendship. Walter Boyajian, for all we keep learning and living together. Thanks to Nenad and Boro Skalonja, Faustino L\'opez, and so many others.
Thank you Eva M. --despite living 10000 km away from each other, you are extremely close to me. Thank you Magda, for showing me the real meaning of the word happiness. Thank you to my parents, Valent\'in and Araceli, for your emotions; for a path of unconditional support. Thank you to my grandmothers, Julia and Inmaculada. I do not underestimate the work you have done to provide the next generations with a better life. I cannot do less now than dedicating this Thesis to you.

}
\clearpage  

\newpage{\
\thispagestyle{empty}}


\addtotoc{Abstract}  
\abstract{
\addtocontents{toc}{\vspace{0em}}  

\begin{flushright}

\textit{``Joy lies in the fight,\\ in the attempt,\\ in the suffering involved,\\ not in the victory itself."}

M. Gandhi

\end{flushright}

In the present work we develop a formalism to tackle the problem of optimal execution when trading market securities. More precisely, we introduce a utility function that balances market impact and timing risk, with this last being modelled as the very negative transaction costs incurred by our order execution. The framework is built upon existing theory on optimal trading strategies, but incorporates characteristics that enable distinctive execution strategies. The formalism is complemented by an analysis of various impact models and different distributional properties of market returns. 

}

\newpage{\
\thispagestyle{empty}}

\clearpage  


\setstretch{1.3}  

\pagestyle{fancy}  

\lhead{\emph{Contents}}  
\tableofcontents  

\newpage{\
\thispagestyle{empty}}

\clearpage

\setstretch{1.5}  
\clearpage  
\lhead{\emph{List of Abbreviations}}  
\listofsymbols{ll}  
{
\textbf{TCA} & \textbf{T}ransaction \textbf{C}ost \textbf{A}nalysis \\
\textbf{TC} & \textbf{T}ransaction \textbf{C}osts \\
\textbf{PnL} & \textbf{P}rofit a\textbf{n}d \textbf{L}oss \\
\textbf{AC} & \textbf{A}lmgren-\textbf{C}hriss \\
\textbf{OTC} & \textbf{O}ver \textbf{T}he \textbf{C}ounter \\
\textbf{RFQ} & \textbf{R}equest \textbf{F}or \textbf{Q}uote \\
\textbf{OG} & \textbf{O}rder \textbf{G}eneration \\
\textbf{PM} & \textbf{P}ortfolio \textbf{M}anager \\
\textbf{IT} & \textbf{I}nformation \textbf{T}echnology \\
\textbf{IS} & \textbf{I}mplementation \textbf{S}hortfall \\
\textbf{IM} & \textbf{I}mpact \textbf{M}odel \\
\textbf{MI} & \textbf{M}arket \textbf{I}mpact \\
\textbf{PMI} & \textbf{P}ermanent \textbf{M}arket \textbf{I}mpact \\
\textbf{TMI} & \textbf{T}emporary \textbf{M}arket \textbf{I}mpact \\
\textbf{VaR} & \textbf{V}alue \textbf{a}t \textbf{R}isk \\
\textbf{UF} & \textbf{U}tility \textbf{F}unction
}
\lhead{\emph{List of Tables}}  
\listoftables  

\lhead{\emph{List of Figures}}  
\listoffigures  

\newpage{\
\thispagestyle{empty}}

\clearpage

\setstretch{1.3}  

\pagestyle{empty}  
\dedicatory{A mis abuelas, Julia e Inmaculada, \\que comenzaron un largo camino para que esta tesis fuera posible}

\addtocontents{toc}{\vspace{1em}}  

\newpage{\
\thispagestyle{empty}}

\mainmatter	  
\pagestyle{fancy}  



\chapter{Introduction} 
\label{Chapter1}
\lhead{Chapter 1. \emph{Introduction}} 

\begin{flushright}

\textit{``A person's work may not be finished in his lifetime, but let us begin."}

J. F. Kenedy

\end{flushright}

%

~

A few years ago, in a scientific conference held at the heart of Europe, I was sitting at a dinner table where an interesting discussion arose. Someone asked the question ``what --in our opinion-- had been the most important discovery or development in human history". To this day, I remember with clarity some of the answers (and the names of the Professors and students who formulated them --although I will omit these here). ``The theory of General Relativity'' --said the Cosmology Professor. ``Black-Scholes' theory'' --said the Mathematical Finance Professor. ``The theory of Evolution'' --said a Biologist. ``The fact that matter is made of atoms'' --said a Condensed Matter Physicist. Fortunately, some students incorporated candidates such as ``the development of mechanical tools" or ``the emergence of agriculture". My answer at the time was ``the development of the scientific method" --the process through which our understanding of nature is hypothesised and tested using logic, reason, and scrutinised by comparison to empirical evidence. Although I am profoundly aware --as the previous examples suggest-- of synthetic happiness and of the power to convince ourselves that what we do is certainly the most relevant and noble task that could possibly be done, perhaps with the aid of this `self-convincing effect', I have, over recent years, come to the belief that ``the existence of markets" stands up there, as one of the greatest developments in the history of mankind. The organisation of societies, not through authoritarian leadership, but through an economic framework which allows individuals to attain Ôunanimity without consentÕ, respecting the fundamental value of individual freedom, and thus reaching a true representative democracy, is in my mind one of the most profound developments of our civilisation.

~

The fact that the existence of a market system might not naturally come to mind when thinking about human progress is best expressed in the words of Kenneth Arrow \cite{Arrow74}: 

{\it ``In everyday, normal experience, there is something of a balance between the amounts of goods and services that some individuals want to supply and the amounts that other, different individuals want to sell. Would-be buyers ordinarily count correctly on being able to carry out their intentions, and would-be sellers do not ordinarily find themselves producing great amounts of goods that they cannot sell. This experience of balance is indeed so widespread that it raises no intellectual disquiet among laymen; they take it so much for granted that they are not disposed to understand the mechanism by which it occurs."}

It is thus important to realise that, although old in nature, markets and trade have not always been around. Furthermore, the realisation that the market system --as opposed to a centralised government-- can enable the organisation of societies \cite{Smith76} is a formidable discovery that deserves special attention. What are the mechanisms that guide this `invisible hand'? How do they operate? Are they --in some sense-- efficient? Does a central agent need to supervise its activity? Is this mechanism preferred to other forms of societal organisations? Remarkably, the answer to these questions is still under intense debate. In the words of F. H. Hahn \cite{Hahn70}: {\it ``The most intellectually exciting question in our subject remains: is it true that the pursuit of private interest produces not chaos but coherence, and if so how is it done?''}. In order to bring some light into the operation of markets, mathematical models have been developed, aiming to address fundamental questions such {\it how prices are formed in a market economy}. From a general perspective, this is the central question around which this Thesis is articulated. From a more specific point of view, the principal question we address is: 
{\it Given how (our own) supply and demand affects prices, what is the ``optimal" way to sell or buy a particular asset?'}.
Specific interest should be put into the keyword ``optimal". What does it mean to sell or buy optimally? What are the ways to address this mathematically? As we will see, how (our own) supply/demand affects prices will be described by {\it market impact models}, while how to ``optimise" our trading will be addressed by the minimisation of a utility function capturing the essential degrees of freedom that contribute to variations of prices or --in our context-- {\it costs}. 

~

Our theory finds its fundamentals in the Walrasian General Equilibrium formulation \cite{Walras74}, in the sense that through the minimisation of a utility function --for a fixed number of shares to buy or sell-- we determine a unique, stable, optimal execution strategy for a trading agent in the market economy. 
Under classical microeconomics theory, the equilibrium price of a financial asset is determined by the point at which supply equals demand. At a fundamental level, the process of price formation therefore depends on the dynamics of an {\it order book}, capturing the evolution of all buy and sell quotes, or in other words, on the {\it market microstructure}, i.e. how the mechanics of trading governs the shape of the supply and demand curves (and hence ultimately the price of the asset). Being a Physicist, I could not avoid seeing some sort of similarity between Statistical Physics and Market Microstructure: in a thermodynamic system particles exchange (units of) energy to reach an equilibrium temperature, while in a financial system traders exchange (units of) the asset (say stock) to reach an equilibrium price. Indeed, I later realised that this specific idea had already been investigated \cite{Foley94}, and that a broad body of literature on price formation and market microstructure has been produced over the last years; see \cite{Friedman62, Ohara95, Harris02, Hasbrouck07, Johnson10, Kissell14, Cartea15, Bouchaud18} to cite a few. In essence, the curiosity to understand the mechanisms that lead to the {\it emergence} of prices in financial markets and the will to deepen my understanding of market microstructure cemented the starting point of this Thesis. At the same time, the use of market prices and traded volumes as signals to economic outputs \cite{Hayek45, Keynes36} indicated the importance of this research within an even bigger picture.
In this work we focus in particular on {\it transaction costs}, i.e. the difference between the price at which we buy or sell an asset, and a specific reference price of that asset (which we will call {\it benchmark price}). 
For simplicity, our discussion will be centered around the Equity market, and so we will talk about buying or selling shares of a publicly-owned company. The study can nonetheless be straightforwardly extended to other asset classes, such as Forex Exchange, Fixed Income, or Commodities. Just as prices depend on supply and demand, transaction costs will depend on the number of shares sold and bought (trading volumes) over a specific time period. Our goal will be focused on the understanding of how our own supply/demand will affect the share price, and given this relationship (impact model), we address the question on what is the best way to buy or sell the stock over time in order to minimise transaction costs. 

~

Over the last two decades, the study and characterisation of transaction costs, generically referred as {\it Transaction Cost Analysis} (TCA), has become an essential component in the investment process. The importance of this field becomes self-evident when observing the difference between the ``on-paper" Profit \& Loss (PnL) and the realised PnL of a portfolio: A theoretically profitable trading strategy can become highly unprofitable when taking into account price moves due to our own supply and demand of the constituent assets. As mentioned before, transaction costs (TC) are defined as follows:
{\it Given a trading order, defined by {\it side} (buy/sell), {\it size} (number of shares), and {\it trading horizon} (time span over which the order must be completely executed), TC are the difference between the average execution price per share and a benchmark price per share,}
\beq \label{TCdef}
c = \xi( p^{\rm exe} - p^{\rm bmk} ),
\eeq
with $\xi = -1$ for a buy order and $\xi = 1$ for a sell order. Here $c$ indicates ``cost'', while $p$ refers to price {\it per share}. The specification ``average'' (when referring to execution price) comes from the fact that trading over a time span can be done by splitting the total number of traded shares over multiple portions (executed at different times and at different prices), in which case the execution price over the trading horizon ought to be calculated as the (volume-weighted) average execution price over all portions. According to the definition (\ref{TCdef}), the value of TC thus highly depends on our choice of benchmark. As we shall see, this can vary in complexity, from a simple market price at a specific point in time, to expected prices determined through sophisticated models, that capture the different factors contributing to price moves. Intuitively, the concept of TC is linked to the difficulty to execute at a target price, due to actual liquidity and market conditions. With this in mind, we can establish three generic categories of TC: 
\begin{enumerate}[i)]
\item {\it Commissions, fees and taxes:} These correspond to extra payments to external parties (intermediaries in the financial transaction), associated to e.g.  brokers/dealers performing the transaction, or to the government allowing it.
\item {\it Spreads:} These correspond to the difference between the ask (offer) and bid price of an asset. Such premioum is associated to market makers bearing the risk of holding the asset available for both sides of the transaction. Therefore we could refer to the spread as the {\it `price of liquidity'}.
\item {\it Price moves in the market:} These correspond to price shifts coming from our own action or inaction in the process of executing an order. While our supply/demand signal causes a change of the asset price {\it (`market impact')}, delaying the execution of an order --with respect to the time it is conceived-- causes the asset price to drift away due to market volatility generated by other market participant's supply and demand {\it (`timing risk')}.
\end{enumerate}

Although, as we mention below, the first two categories can be easily incorporated into our framework, we will focus here on TC associated to our own action (or inaction) with respect to trading an order. In this context, the fundamental problem of TCA can be formulated in terms of the {\it trader's dilemma}:

{\it If an order is executed quickly, this may move the security's price against the trader {\it (`market impact')}; if, however, the order is executed slowly, market volatility might lead to a less favourable price that when the order was received by the trader {\it (`timing risk')}.}

This concept is best captured in the words of A. F. Perold \cite{Perold98}: {\it ``Reality involves the cost of trading {\it and} the cost of {\it not} trading''}, and conveys the idea that there must be an optimal rate of execution, determined by optimising transaction costs. Here the terms ``quickly'', ``slowly'', and ``optimal rate of execution'' are precisely defined by the balance between market impact and timing risk. 

~

This conceptual optimisation problem was captured more precisely and mathematically modelled by Almgren and Chriss (AC) \cite{AlmgrenChriss01}, and the goal of this Thesis is to build upon this work in order to develop an {\it alternative formulation of timing risk}. Here we introduce a new utility function, characterising timing risk beyond ``variance of transaction costs". Specifically, we consider timing risk as the (negative) tail of the TC distribution, and in this way go beyond a ``mean-variance" approach to TCA. As we will see, solving for the optimal trading strategy under this new approach allows us to obtain strategies with more shares traded in later trading intervals than in earlier intervals (a characteristic not possible under the AC framework with a linear impact model). Furthermore, we study the behaviour of the optimal trading strategy when instead of normally-distributed returns (as AC assume), one considers market returns following a Student distribution. As it will be argued, this is a more accurate characterisation of market returns due to the fat-tail nature of their distribution. We also investigate different impact models, such a sub-linear dependence of the stock price with the number of executed shares, and an exponentially or power-law time dependence of market impact in a propagator model \cite{Bouchaud18}. By doing all this we determine some of the characteristics and limitations of the AC formalism, as well as of our own. In particular, we find that as we consider timing risk encapsulating more extreme market events, the optimal trading strategy ``becomes more impatient'' (in the sense that most of the shares ought to be executed close to the start of the order). Importantly, we also find that if we minimise a utility function containing only our timing risk term, the optimal trading path is such that all shares ought to be executed at the order start. Finally we discuss the extension of our formalism to include other degrees of freedom, such as the treatment of the trading horizon as a variable --rather than a parameter-- of the problem, and offer other future prospects of our work.

~

~

~

~

~

~

~

~

~

~

~

~

~

~

~

~

~

~

~

~


\chapter{Trading Strategies\\ and Transaction Costs} 
\label{Chapter2}
\lhead{Chapter 2. \emph{Trading Strategies and Transaction Costs}} 

\begin{flushright}

\textit{``It is not from the benevolence of the butcher, \\
the brewer, or the baker, that we expect our dinner, \\
but from their regard to their own interest."}

Adam Smith

\end{flushright}

\section{The investment process}

As it was described in the introduction, in this Thesis we tackle the problem of minimising transaction costs, defined by equation (\ref{TCdef}), and we focus on the aspect of transaction costs arising from price moves in the market. In order to solve this optimisation problem, it is useful to break down the execution of an order as it occurs over time. An {\it order} to trade a particular security is an instruction specified by the following parameters:
\begin{itemize}
\item {\it Side} $\xi$: buy or sell.
\item {\it Size} $N$: total number of shares to trade.
\item {\it Time horizon} or {\it trading horizon} $T$: time given to complete the order.
\end{itemize}
As we will see, there is a large variety of trading algorithms, which depend on market characteristics such as traded volumes, volatilities and price levels.

\newpage

We will here classify orders into four categories:
\begin{enumerate}[1)]
\item {\it Market order:} This is an order to be executed during the time horizon. As such, their execution is typically benchmarked with respect to ``expected" prices given by specific trading algorithms.
\item {\it Market-on-Close order:} This is an order to be executed at the close of the Exchange. As such, their execution is typically benchmarked with respect to the price at the time of the close.
\item {\it Market-Open order:} This is an order to be executed at the open of the Exchange. As such, their execution is typically benchmarked with respect to the price at the time of the open.
\item{\it Limit order:} This is an order whose execution should only be carried during times at which the share price is between a band of threshold prices. As such, their execution should be typically benchmarked with respect to prices in which it is possible (as allowed by this band) to execute the order.
\end{enumerate}

In these definitions, we have assumed that orders are done through Exchanges (where a {\it limit order book} (LOB) exists). Although we will focus on such scenario here, it is important to keep in mind that this is not always the case. Often, trading is done {\it ``over-the-counter"} (OTC), where trading is carried through a specific dealer. This distinction is relevant because for OTC trading we lack ``tick data", i.e. a quasi-continuum of prices through the trading horizon, implying that the execution algorithms discussed here do not generally apply in this case. Furthermore, OTC execution is typically benchmarked by comparing various quotes from different dealers. In reality, {\it Request-For-Quote} (RFQ)-based trading is more popular in Fixed Income markets, while LOB-based trading is more standard in the Equity market.

~

\begin{figure}
  \begin{center}
    \includegraphics[width=\textwidth]{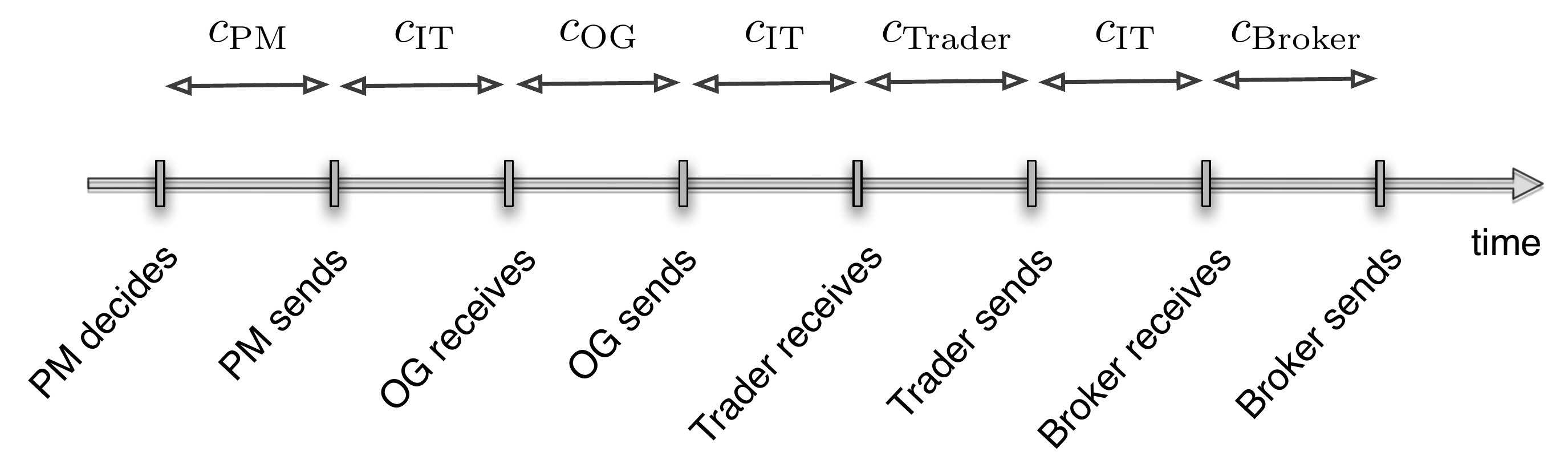}
  \end{center} 
\caption[Timeline of the investment process.]{Timeline of the investment process. The order, decided originally by the Portfolio Manager (PM), is sent to Order Generation (OG) and subsequently to Trader and Broker. Each of the intervals between the time at which the order is received and the order is sent has an associated cost. Furthermore, each of the time intervals between when the order is sent and the order is received has an extra cost, which we label as `Information Technology' (IT) cost.}
  \label{FigInvestmentProcess}
\end{figure}

In figure \ref{FigInvestmentProcess} we schematically show the different stages of an order in a typical trading floor. An order is first decided and then sent to an Order Generation (OG) team (or system) by the Portfolio Manager (PM). Such OG team collects orders from multiple PMs and sends them to the Trading team. The trader can then evaluate, given the trading horizon, how to partition and execute the order. Typically, larger order volumes will be executed in time periods with larger market liquidity. The trader could at this stage put in place his strategy by executing through an Exchange. In practice, however, the order is sent (or `placed') to a Broker with a series of instructions (urgency level, etc.), who will in turn implement the execution strategy by trading through an Exchange. Finally, relevant execution values (fill times, fill volumes, and fill prices) are retrieved from the Broker\footnote{It is worth noticing that, in reality, different PMs might send orders referring to the same stock to OG. In this case, these orders are considered as {\it `child orders'}, and blocked (typically by the Trader or by OG) into a single {`parent order'}. In this case, TCA for Traders and Brokers ought to be done with respect to the parent order (since after blocking, the information about the individual child orders is lost). It is also important to notice that after the order execution information is retrieved from the Broker, an `Order Allocation' team (or Traders themselves) allocate the position and PnL to the different portfolio accounts involved in the trade.}. To each of these stages we can associate a transaction cost: $c_{\rm PM}$, $c_{\rm OG}$, $c_{\rm Trader}$, $c_{\rm Broker}$, which allows us to better identify where improvements should be made in order to minimise TC.

\section{Transaction costs}

Given a trading order, an {\it execution algorithm} is defined as the path $\{n(t)\}$, $0\leq t \leq T$, describing the number of shares being executed at time $t$. Here $T$ is the order time horizon, and $t=0$ is the {\it `order start'} time. Depending on when we consider that the order becomes ``active", the latter can correspond to the time at which the Trader receives the order (typical scenario), the time at which the Broker or OG receive the order, or to the time at which the PM decides that the order should be traded. If $N$ is the total number of shares to buy or sell, $n(t)$ must satisfy the constraint $ \int_{0}^T n(t) dt = N $. In reality, however, execution is a discrete process: it occurs as a series of $F$ events (fills, placements, ...) at a set of times $\{t_{j=1,\ldots,F}\}$, and thus $n(t) = \sum_{j=1}^F n_j \delta(t-t_j)$, being $\delta(t)$ a Dirac delta function. Similarly, execution can be thought as taking place over K time sub-intervals of time-length $\Delta t$ in which $[0,T]$ is subdivided, and in each of these sub-intervals $n_k^{\rm exe}$ shares being executed at price $p_k^{\rm exe}$, where $k =1,\ldots,K$. This leads to the constraint
\beq \label{TCAconstraint}
\sum_{k=1}^K n^{\rm exe}_k = N,
\eeq
with $K = T/\Delta t$. We will then call {\it trading strategy}, {\it execution strategy}, {\it trading algorithm}, or {\it execution algorithm} to the path $\{n_k^{\rm exe}\}$, with $k=1,\ldots,K$ (we will denote $n_k^{\rm exe} = n_k$ and $p_k^{\rm exe} = p_k$ indistinctly throughout the text).
It is important to notice that a trading strategy $\{n_k^{\rm exe}\}$ can itself be employed as a benchmark (in which case it will be denoted as $\{n_k^{\rm bmk}\}$). This can be done from two different perspectives:
\begin{enumerate}[i)]
\item {\it Pre-trade:} $\{n_k^{\rm bmk}\}$ and $\{p_k^{\rm bmk}\}$ depend on values prices previous to trade execution.
\item {\it Post-trade:} $\{n_k^{\rm bmk}\}$ and $\{p_k^{\rm bmk}\}$ depend on values prices posterior to trade execution.
\end{enumerate}

With the discretisation of the time horizon into $K$ subintervals, we can write
\beq\label{p_exe_p_bmk}
p^{\rm exe} \equiv \frac{1}{N} \sum_{k=1}^K n_k^{\rm exe} p_k^{\rm exe}, \quad\quad
p^{\rm bmk} \equiv \frac{1}{N} \sum_{k=1}^K n_k^{\rm bmk} p_k^{\rm bmk}
\eeq
Here
\beq
N &\equiv& \mbox{number of shares to buy/sell within time}\; T. \nonumber\\
n_k^{\rm exe} &\equiv& \mbox{number of executed shares in time bin}\; k. \nonumber\\
p_k^{\rm exe} &\equiv& \mbox{average execution price in time bin}\; k. \nonumber\\
n_k^{\rm bmk} &\equiv& \mbox{number of shares in time bin}\; k \; \mbox{executed by the algorithm/benchmark.} \nonumber\\
p_k^{\rm bmk} &\equiv& \mbox{average algorithm/benchmark price in time bin}\; k. \nonumber
\eeq 
Combining equations (\ref{TCdef}) and (\ref{p_exe_p_bmk}), we can write {\it transaction costs} as\footnote{Note that if we wish to express TC in bips, this definition should be written as $c(bps) = 10^4 \frac{p^{\rm exe}-p^{\rm bmk}}{p^{\rm bmk}}$.\\ It is worth noticing that this definition can be embedded within a most generic one. In the continuum time limit, we can write transaction costs per share as
\beq\label{TransactionCostsEq}
c \equiv \int_0^T \Lambda(t) dt,
\eeq
where $\Lambda(t) \equiv \sum_{k=1}^K \kappa(t) \delta(t-k\Delta t)$ is a cost function whose functional form depends on our precise definition of transaction costs; in our case $\kappa(t) \equiv \frac{\xi}{N} \left( n^{\rm exe}(t) p^{\rm exe}(t) - n^{\rm bmk}(t) p^{\rm bmk}(t) \right)$ (linear function of the number of shares and price).}
\beq\label{TCdefLong}
c \equiv \frac{\xi}{N} \sum_{k=1}^K \left( n_k^{\rm exe} p_k^{\rm exe} - n_k^{\rm bmk} p_k^{\rm bmk} \right)
\eeq
Where $\xi = -1$ for a buy order and $\xi=1$ for a shell order\footnote{If within a trading bin $k$ we have $F$ fills/placements, the average execution and benchmark prices in that bin are calculated, respectively, as $p_k^{\rm exe} = \frac{1}{F} \sum_{j=1}^F n_j^{\rm exe} p_j^{\rm exe}$, $p_k^{\rm bmk} = \frac{1}{F} \sum_{j=1}^F n_j^{\rm bmk} p_j^{\rm bmk}$.}. Transaction costs are therefore characterised by $\{n_k^{\rm bmk}\}$ and $\{p_k^{\rm bmk}\}$.  In Appendix~\ref{AppBenchmarks} we specify the functional form of these variables for the most standard benchmarks and trading algorithms. In this Thesis we focus in particular in the {\it implementation shortfall} (IS) definition of transaction costs, i.e. henceforth we will consider
\beq \label{EqTCIS}
c \equiv \xi \left[ \frac{1}{N} \sum_{k=1}^K n_k^{\rm exe} p_k^{\rm exe} - p_0 \right],
\eeq
being $p_0$ the stock price at the order start.

\begin{figure}[t]
\begin{center}
\includegraphics[width=0.77\linewidth]{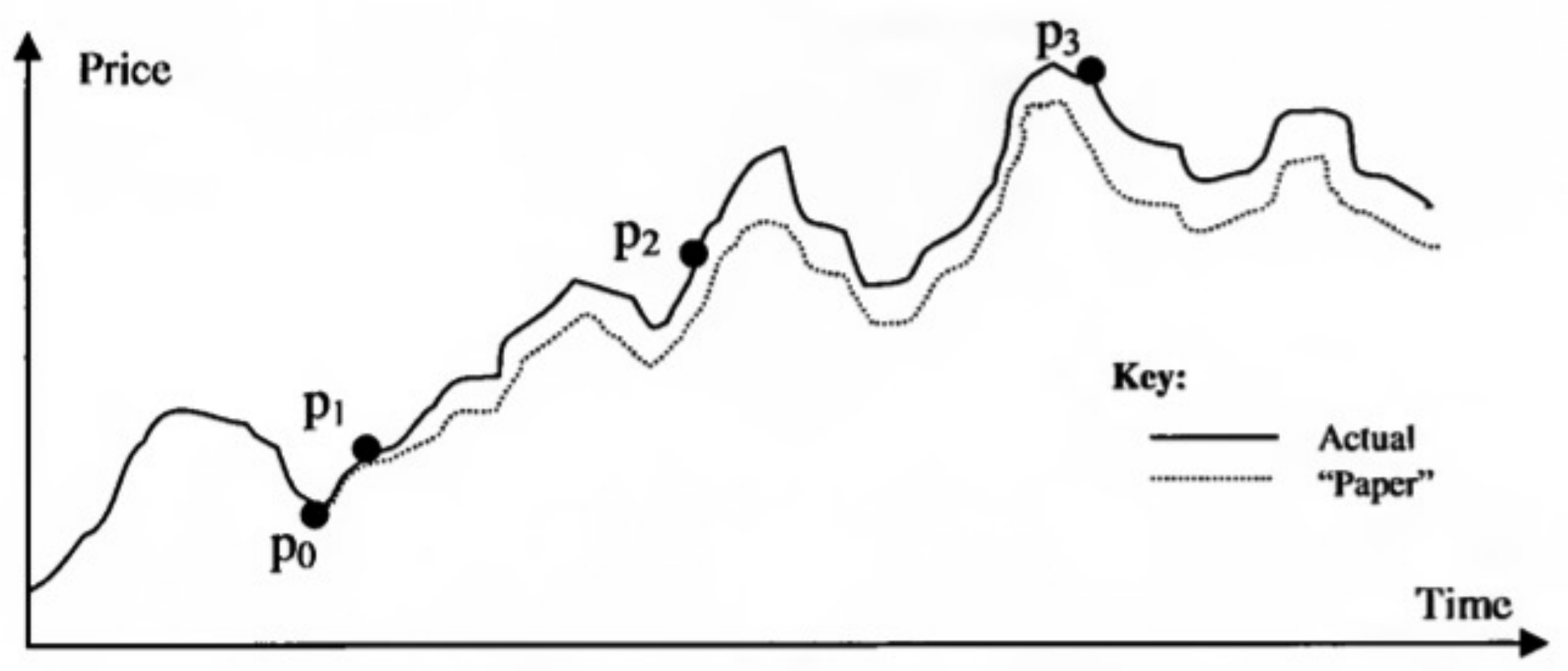}
\caption[Pictorial representation of market impact.]{Pictorial representation of market impact. The black line represents the stock price in the presence of our trading (``Actual"), while the grey dotted line represents the stock price in the absence of our trading (``Paper"). Since either we do execute the order or we do not, we cannot observe both components, and thus market impact is not a directly observable quantity (its magnitude can only be inferred via a {\it calibrated} impact model). $p_0$ denotes the price at the order start (in this case when the market receives information of our interest in the security). $p_1$ and $p_2$ represent the market prices at fill times. $p_3$ indicates the market price at the time from which the security's price in the absence and in presence of our trade differ by a fixed amount {\it (`permanent market impact')}. Before this time, these two prices differ by an amount that may vary with time {\it (temporary market impact)}. Figure taken from \cite{Johnson10}.}
\label{MarketImpactFig}
\end{center}
\end{figure}

We next define some of the main terms used in TCA. To this end, it is useful to introduce a simple price model. Consider that when trading $n_\tau$ shares in the time interval $[0,\tau]$ the stock price moves as\footnote{A small variation of this simple model is to consider $p_\tau - p_0 = {\cal I}(n_\tau) + \zeta_\tau$, i.e. a price model for arithmetic returns. Notice that there is no assumption of preferred market directionality, i.e. not drift or momentum other than the one induced by our own trading.}
\beq\label{SimpleImpactModel}
\frac{p_\tau - p_0}{p_0} = {\cal I}(n_\tau) + \zeta_\tau
\eeq
Here ${\cal I}(n_\tau)$ is a (deterministic) function of the number of executed shares, while $\zeta_\tau$ is a stochastic component with zero mean, $\mathbb{E}[\zeta_\tau]=0$, capturing the market volatility from $t=0$ to $t=\tau$. It is custom to use a model in which $\zeta_\tau = \sigma \sqrt{\tau} \chi_\tau$, with $\chi_\tau$ having zero mean and standard deviation 1, but is not necessary to particularise the specific form of $\zeta_\tau$ at this stage. Taking the expectation at $t=0$ of equation (\ref{SimpleImpactModel}) we obtain
\beq
\frac{\mathbb{E}_0[p_\tau] - p_0}{p_0} = {\cal I}(n_\tau)
\eeq
This quantity will be referred as {\it `market impact'} or {\it `(market) impact function'}. Here we have specified the average at $t=0$, time at which we {\it predict} a particular market impact based on the size of our order and the functional form of ${\cal I}(n_\tau)$. It is important to realise that market impact can be {\it estimated} (what we believe what it was) or {\it predicted} (what we believe it will be), but {\it market impact can never be measured}. As illustrated in figure \ref{MarketImpactFig}, market impact is the difference between the stock price in the presence (``Actual") and in the absence (``Paper") of our execution. Since in a particular market realisation, we either execute or do not execute, we can only observe one, not both, of these components. This is why {\it impact models} (explained in detail in section \ref{SecImpactModels}) are one of the most, if not the most, relevant elements of TCA: they allow us to estimate (or predict) the difference between the ``Actual" and ``Paper" stock price.

~

Given the simple model and the equations above, we can now give definitions to some of the most recurring concepts in TCA\footnote{In the literature, and in the industry, different terms are used to describe the same (or similar) quantities. For example, {\bf `execution cost'} or {\bf `trading cost'} are synonyms of `transaction cost'. {\bf `Expected impact cost'} or {\bf `impact cost'} describe `expected transaction cost', while `timing risk' is also referred as {\bf `timing cost'}, {\bf `shortfall risk'}, {\bf `volatility risk'}, or {\bf `opportunity cost'}. Furthermore, {\bf `implementation shortfall'} and {\bf `slippage'} are mentioned as similar concepts to `market impact' (which some authors define, not as an expectation value, but as a stochastic variable, including a noise term). We should notice that some of these definitions differ from those given by Perold \cite{Perold98}, and given in this text. One important example is that of \textbf{\emph{`opportunity cost'}}, defined as the appreciation (in currency value) of the shares specified by the order but not executed by the end of the trading horizon. If $\tilde{N}$ is the number of shares specified by the order and $N$ is the number of shares actually executed by the end of the time horizon, the opportunity cost is then given by
\beq
\xi \; \frac{\tilde{N}-N}{N} \; \Delta p_T,
\eeq
where $\Delta p_T \equiv p_T - p_0$ is the security's market price difference over the time interval $[0,T]$ (and which needs to be estimated/predicted via an impact model), and $\xi = -1$ for a buy order, $\xi=1$ for a sell order. Notice that in the text we take $N = \tilde{N}$, i.e. we do not discuss the case in which the order is partly executed by the end of the trading horizon, and thus --given this definition-- opportunity cost is zero. The case in which $N<\tilde{N}$ (non-zero opportunity cost) is then left as a future prospect of this work.
}:
\begin{itemize}
 
\item \textbf{Transaction cost:} $c$. This is the (signed) difference between average execution price and benchmark price (c.f. equations (\ref{TCdef}), (\ref{TCdefLong})). The sign included in this definition implies that we take positive costs as out-performance with respect to the benchmark, and negative costs as under-performance with respect to the benchmark. It is important to notice that, while in post-trade analysis transaction cost is a deterministic variable, from a pre-trade perspective it is a stochastic variable. This is why we explicitly define `expected transaction cost' next.
 
\item \textbf{Expected transaction cost:} $\mathbb{E}[c]$. This is the expectation of transaction cost. From a post-trade perspective, both expected transaction cost and transaction cost are interchangeable concepts. From a pre-trade perspective, however, transaction cost is a stochastic variable, and expected transaction cost refers to its expectation value.
 
\item \textbf{Market impact:} $\frac{\mathbb{E}[p_t] - p_0}{p_0}$. Calculated at a specific time $t$, this is the expected transaction cost when the benchmark is taken to be the security's price at the order-start time, $p_0$. The expectatino value is also taken at the order start time (i.e. $\mathbb{E}[c] \equiv \mathbb{E}_0[c]$). Notice that market impact can be defined in bips (as done here) or, equivalently, in currency value, in which case ${\rm market \; impact} \equiv \mathbb{E}[p_t] - p_0$.
 
\item \textbf{Timing risk:} $\frac{\sigma[c]}{p_0}$. This corresponds to the standard deviation of transaction cost, i.e. the square root of the variance $\mathbb{V}[c]$. Notice that timing risk can be defined in bips (as done here) or, equivalently, in currency value, in which case ${\rm timing \; risk} \equiv \sigma[c]$.

\end{itemize}

\section{Impact models} \label{SecImpactModels}

An {\it impact model} (IM) postulates a certain relationship between the number of shares executed over a time interval and the change in the security's price. In terms of equation (\ref{SimpleImpactModel}), the impact model corresponds to the functional form assumed for ${\cal I}(n_\tau)$. In the following, we first explain the IM proposed by Almgren and Chriss \cite{AlmgrenChriss01}, and then introduce our own impact model, to finally briefly comment on other models in the literature.

Almgren and Chriss split market impact (MI) into the sum of two components:
\begin{enumerate}[i)]
\item {\it Permanent market impact (PMI):} This is the difference between the stock price in the presence and in the absence of execution when this value becomes stationary, i.e. when it does not change with time. In terms of the model, PMI has a functional form non-dependent on time, and whose effect on the stock price is delayed by one time sub-interval.
\item {\it Temporary market impact (TMI):} This is the difference between the stock price in the presence and in the absence of execution before this value becomes stationary, i.e. when it still changes with time. In terms of the model, TMI has a functional form dependent on time, and whose effect on the stock price is not delayed in time.
\end{enumerate}

More specifically, if the trading horizon $[0,T]$ is subdivided into $K$ subintervals of length $\Delta t=T/K$ each, AC assume that, when we execute $n_k$ shares in the sub-interval $k$, the stock price changes as
\beq \label{pkAC}
p_k = p_0 + \sum_{j=1}^{k-1} \Big[ \sigma \sqrt{\Delta t} \; \chi_{j} \underbrace{- \Delta t \; g\left(\frac{n_{j}}{\Delta t}\right)}_{\rm PMI} \Big] \underbrace{- h\left( \frac{n_k}{\Delta t}  \right)}_{\rm TMI} \\
\nonumber\\
\nonumber\\
\nonumber
\eeq
Here $\chi \sim {\cal N}(0,1)$ (normally-distributed variable with mean 0 and standard deviation 1), $p_0$ is the stock price at order start, and $g(n)$, $h(n)$, are {\it permanent} and {\it temporary} impact functions, respectively, which are assumed to be of the form\footnote{While the original work of AC assumes a linear dependence of the stock price with the rate of execution, $n_\tau/\tau$, in a subsequent work \cite{Almgren03} Almgren extends this theory to sub-linear models of TMI. This, and models alike, are currently widely used in the industry \cite{Almgren05}.}.
\beq
g(x) &=& \xi \; \gamma \; x \label{ACPMI} \\
h(x) &=& \xi \; ( \varepsilon + \eta \; x ) \label{ACTMI}
\eeq
being $\gamma$, $\epsilon$ and $\eta$ model parameters to be calibrated according to historical data. As we will see, this linearity of both PMI and TMI with the number of executed shares is an essential assumption that defines the characteristics of the optimal execution given by AC. Actually, the linear relationship between the number of executed shares and the stock price in the PMI component is rather generic: Huberman and Stanzl proved that this is the only functional form for PMI to be compatible with non-arbitrage conditions\footnote{For a comprehensive discussion of the functional forms of market impact (under a propagator framework), compatible with a non-arbitrage assumption, see \cite{Gatheral09}.} \cite{HubermanStanzl04}. In these equations, $\varepsilon$ can be interpreted as a bid-ask spread cost. Assuming that we execute at the price (\ref{pkAC}), and inserting this expression into equation (\ref{EqTCIS}), we obtain
\beq \label{ACcostDerivation}
c &=& \xi \left[ \frac{1}{N} \sum_{k=1}^N n_k \Big\{ p_0 + \sum_{j=1}^{k-1} \left[ \sigma \sqrt{\Delta t} \; \chi_{j} - \Delta t \; g\left( \frac{n_{j}}{\Delta t} \right) \right] \Big\} - \frac{1}{N} \sum_{k=1}^N n_k h\left( \frac{n_k}{\Delta t}  \right) - p_0 \right] \nonumber\\
&=& \frac{\xi}{N} \left[ \sum_{k=1}^N \sum_{j=0}^{k-1} \Big\{ n_k \left[ \sigma \sqrt{\Delta t} \; \chi_{j} - \Delta t \; g\left( \frac{n_{j}}{\Delta t} \right) \right] \Big\} - \sum_{k=1}^N n_k h\left( \frac{n_k}{\Delta t}  \right) \right]
\nonumber\\
&=& \frac{\xi}{N} \left[ \sum_{j=0}^{N-1} \sum_{k=j+1}^{N} \Big\{ n_k \left[ \sigma \sqrt{\Delta t} \; \chi_{j} - \Delta t \; g\left( \frac{n_{j}}{\Delta t} \right) \right] \Big\} - \sum_{k=1}^N n_k h\left( \frac{n_k}{\Delta t}  \right) \right],
\eeq
where we have used $\sum_{k=1}^N n_k = N$, and $n_0 = 0$, $\chi_0 \equiv 0$. Defining 
\beq 
x_j \equiv \sum_{k=j+1}^N n_k = N - \sum_{k=1}^j n_k,
\eeq
which represents the number of shares not yet executed by the end of interval $j$, and taking into account that $x_N = 0$, we finally get\footnote{It is instructive to remark that should we have considered a MI ``local in time", i.e. a price model $p_k = p_{0} + \sum_{j=1}^k \Big[ \sigma\sqrt{\Delta t} \; \chi_{j} - \Delta t \; g\left( \frac{n_{j}}{\Delta t} \right) \Big] - h\left( \frac{n_k}{\Delta t}  \right)$, we would have obtained the IS transaction cost $c = \frac{\xi}{N} \left[ \sum_{k=1}^K \Big\{ \sigma \sqrt{\Delta t} \; \chi_k - \Delta t \; g\left( \frac{n_k}{\Delta t} \right) \Big\} x_{k-1} - \sum_{k=1}^N n_k h\left( \frac{n_k}{\Delta t}  \right) \right]$, which fundamentally affects some of the conclusions derived from the model. The same would occur if consider a sub-linear TMI model, e.g. a relationship of the type $h(x) = \xi \; ( \varepsilon + \eta \; x^\alpha )$, with $\alpha<1$.}
\beq \label{TCAC}
c^{{}_{\rm AC}} = \frac{\xi}{N} \left[ \sum_{k=1}^K \Big\{ \sigma \sqrt{\Delta t} \; \chi_k - \Delta t \; g\left( \frac{n_k}{\Delta t} \right) \Big\} x_k - \sum_{k=1}^K n_k h\left( \frac{n_k}{\Delta t}  \right) \right]
\eeq
This is the expression for TC that will be used when we calculate the optimal trading strategy given by Almgren-Chriss.

~

We now consider an alternative impact model. Extending the price dynamics given in equation (\ref{SimpleImpactModel}) over multiple time intervals, we have the ``law of motion" for the price:
\beq \label{DMimpactmodel}
p_{k} = p_{k-1} ( 1 + {\cal I}(n_k) + \zeta_k ),
\eeq
where ${\cal I}(n_k)$ is an impact function and $\zeta_k$ a noise term. Assuming that we execute at this price, equation (\ref{EqTCIS}) gives in this case
\beq \label{TCIMDM}
c = \xi \; p_0 \left[ \frac{1}{N} \sum_{k=1}^K \Big\{ n_k^{\rm exe} \prod_{j=1}^{k} \left( 1 + {\cal I}(n_j) + \zeta_j \right) \Big\} - 1 \right]
\eeq
In order to provide a comparison with the model developed by Almgren and Chriss, we will use this expression of TC --assuming various functional forms for ${\cal I}(x)$-- in Chapter~\ref{Chapter3}. Other distinct models exist in the vast body of literature on market impact. To cite a few, the pioneer work of Kyle \cite{Kyle85} postulated a linear function for MI, while later works have shown that the ``experimental" shape of MI is actually concave. For example, J.-P. Bouchaud {\it et al.} postulate a square-root behaviour \cite{Toth11}, while Lillo, Farmer, and collaborators, argue in favour of a logarithmic MI function \cite{Lillo15}. Recent models develop a ``crossover model" for MI, in which, for small executed volumes $n^{\rm exe}$, the stock price is linear with $n^{\rm exe}$, while, for large executed volumes, the stock price is proportional to $\sqrt{n^{\rm exe}}$ \cite{Bucci19}. Other recent studies on MI brilliantly point out that the relevant variable for market impact is the {\it `execution horizon'}, i.e. the time that we take to fully execute the order, rather than the executed volume or the rate of execution \cite{Cont19}. As a future prospect of our work, it will be interesting to apply these models and ideas to the utility function proposed in Chapter~\ref{Chapter3}.

\section{Transaction-cost models}

As explained above, our goal is to determine the execution strategy $\{n_k^{\rm exe} \}$ that minimises transaction costs, caused by {\it price moves in the market}\footnote{Including commission costs can be done by adding a (calibrated/measured) constant to the TC discussed here. Spread costs can be included in the model in various ways. The simplest method is to consider them as a calibrated constant (i.e. as commissions) in the market impact function. More elaborate models of spread costs consider a relationship spread--number of executed shares, and the estimated/predicted spread being added to the estimated/predicted execution price.}, and determined by equation (\ref{EqTCIS}). We should notice, however, that --from a pre-trade perspective-- the price is a stochastic variable, so in this case $c$ is a stochastic variable as well. Then, when we talk about ``minimising transaction costs", what do we mean? One option would be to minimise the expectation value of (\ref{EqTCIS}), subject to the constraint (\ref{TCAconstraint}). Bertsimas and Lo \cite{Bertsimas98} take this approach from a {\it dynamic optimisation} perspective, i.e. they solve recursively a Bellman equation, ensuring that the {\it utility function} $U[c] = -\mathbb{E}[c]$ is minimised at the beginning of every time interval in which the trading horizon is subdivided. With an IM similar to (\ref{DMimpactmodel}) (but with arithmetic --not geometric-- returns) and  with an impact function of the form ${\cal I}(n_k) = \gamma \; n_k$, they obtain that the optimal trading path corresponding to this utility function is a TWAP strategy\footnote{Using more complex impact models they obtain trading strategies with a richer structure.} (c.f. Appendix~\ref{AppBenchmarks}). Almgren and Chriss \cite{AlmgrenChriss01} take, however, a different approach. They realise that by introducing a variance term in the utility function, one may capture the interplay between market impact --``the cost of trading immediately" (represented by $\mathbb{E}[c]$)-- and timing risk --``the cost of {\it not} trading immediately" (represented by $\mathbb{V}[c]$). 
The AC {\it utility function} is\footnote{AC write $U^{{}_{\rm AC}}[c] = - \mathbb{E}[c] + \tilde{\lambda} \mathbb{V}[c]$, which is equivalent to (\ref{ACutility}) under the definition $\tilde{\lambda} \equiv \frac{\lambda}{1-\lambda}$.}
\beq \label{ACutility}
U^{{}_{\rm AC}}[c] = -(1-\lambda) \; \mathbb{E}[c] + \lambda \; \mathbb{V}[c],
\eeq
which they minimise with $c$ given by (\ref{TCAC}), subject to the constraint (\ref{TCAconstraint}), and with the IM defined by equations (\ref{pkAC}), (\ref{ACPMI}), (\ref{ACTMI}). In contrast to Bertsimas and Lo, AC solve a {\it static optimisation} problem, minimising $U^{{}_{\rm AC}}[c]$ at $t=0$ (order start). Therefore, the expectation and variance in equation (\ref{ACutility}) are to be understood at $t=0$. In this equation, $\lambda$ is the {\it `risk-aversion'} parameter, whose magnitude is determined by the ``aggressiveness" of the order ($\lambda \to 0 \Rightarrow$ non-urgent order, $\lambda \to 1 \Rightarrow$ very-urgent order). 

The utility function (\ref{ACutility}) can be therefore interpreted as follows: if we trade aggressively, our sudden demand/supply may move the security's price substantially, thus creating a market impact, which is associated with the term $\mathbb{E}[c]$. In contrast, if we trade passively, price volatility may cause an increase in $\mathbb{V}[c]$, or timing risk. In this way, AC captures mathematically the ``trader's dilemma", described in Chapter~\ref{Chapter1}. As a further remark, one may wonder what does $\mathbb{V}[c]$ have to do with the cost associated to ``not trading immediately" (or timing risk). Taking the standard deviation of the AC transaction cost up to time $t$, we get a timing risk proportional to $\sqrt{t}$, which means that when penalising the variance term in the utility function, we are essentially penalising the lapse of time without completing the order. From this point of view, there is an optimal {\it rate of execution}, that achieves a compromise between market impact and timing risk. This gives us the `optimal trading strategy', which, as we vary the risk-aversion parameter $\lambda$, defines an {\it `efficient frontier'} (set of optimal solutions) in the market-impact/timing-risk two-dimensional space. We can then summarise the AC problem of TCA as follows:

{\it Given a trading order, defined by side, size, and time horizon, and given an impact model of the form described by equations (\ref{pkAC}), (\ref{ACPMI}), (\ref{ACTMI}), find a trading strategy $\{n_k^{\rm exe}\}$, fulfilling the constraint (\ref{TCAconstraint}), that minimises the utility function (\ref{ACutility}), for a given value of  the risk-aversion parameter $\lambda$.}

This model, and variations of it, has become paramount in the financial industry. The optimal strategy can be used --from a post-trade perspective-- as a benchmark for Best Execution reporting and for Trading Performance analysis. From a pre-trade perspective, the optimal strategy is widely used as a trading algorithm. The model by AC has been refined by e.g. considering sub-linear impact models. For example, Almgren {\it et al.} \cite{Almgren05} construct a model with both PMI and TMI given by a power law, with respective calibrated exponents $\alpha \approx 0.9$ and $\beta \approx 0.6$, something that is in approximate agreement with the value $\alpha = 1$ derived from the non-arbitrage arguments of Huberman and W. Stanzl \cite{HubermanStanzl04}, and with the value $\beta = 0.5$ derived from the theoretical considerations of Barra \cite{Barra97}. Other extensions of the AC framework consider additional factors (such as momentum, market volume, spread, or volatility) to contribute to the price move. As such, to the AC optimal cost estimation/prediction, they add a multi-factor model including these variables, and in this way determine a refined estimation/prediction of transaction costs. In the next chapters we will present alternative refinements to the AC formalism. First and foremost, we introduce an alternative utility function, based on a conception of timing risk beyond the variance of TC; second, we will consider non-Gaussian returns in the price evolution; and finally, we will consider TC as given by equation (\ref{TCIMDM}).

~

~

~

~

~

~

~

~

~

~

~

~

~

~

~

~

~

~

~

~

~

~

~

~


\chapter{Execution Costs\\ Beyond Mean-Variance} 
\label{Chapter3}
\lhead{Chapter 3. \emph{Execution Costs Beyond Mean-Variance}} 

\begin{flushright}

\textit{``All models are wrong, but some are useful."}

G. E. P. Box

\end{flushright}

\section{Probability distribution of transaction costs} \label{ProbdistSec}

Given an impact model and a law of motion for the execution price, transaction costs --defined by equation (\ref{EqTCIS})-- follow a specific probability distribution. The optimal execution strategy derived from the model of Almgren and Chriss is such that it balances the mean and the variance of this distribution. However, a natural question may arise: How would the optimal execution strategy change when we take into account further moments of the distribution, or rather, the full probability distribution? This question is particularly important in view that in most financial markets, returns are fundamentally non-Gaussian, or more specifically, they feature fat-tail characteristics \cite{Bouchaud09}. In particular, as we know from portfolio theory, we can think about risk ``beyond the variance of the distribution", for example considering the `Value-at-Risk' (VaR) associated to the probability distribution \cite{Jorion06,Alexander09}. Although there is controversy and criticism on whether VaR is an appropriate measure of overall risk (and whether it can be appropriately estimated/predicted) \cite{Taleb09}, it captures features that the variance of the distribution does not capture (and vice versa!), being particularly suited to characterise extreme events. It becomes thus natural to develop a framework in which the optimal trading strategy is based to minimising both, market impact, as well as (extreme) losses. In this work we will develop such formalism.

\newpage

\begin{figure}
\begin{center}
\includegraphics[width=\linewidth]{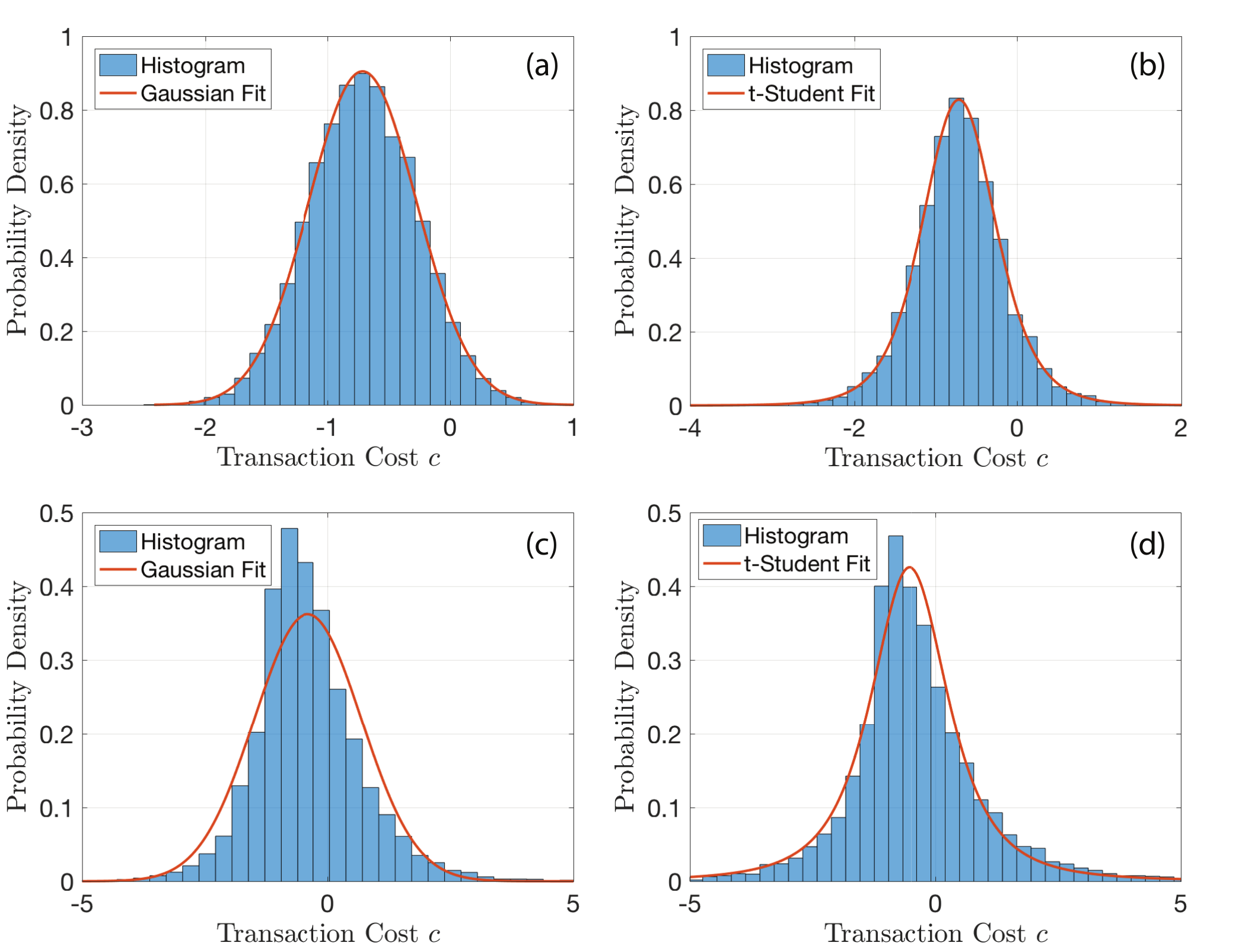}
\caption[Probability density functions of transaction costs.]{Probability density functions of transaction costs. In order to generate the histograms we have generated $10000$ price paths. (a) Histogram and Gaussian fit corresponding to the scenario 1) described in the main text. Here the TC distribution is well described by a normal distribution. The sample of TC has mean $= -0.71$, standard deviation $=0.44$, skewness $=0.041$, and kurtosis $=3.0$. The Gaussian fit has mean $=-0.71\pm 0.01$ and standard deviation $=0.44\pm 0.01$. (b) Histogram and t-Student fit corresponding to the scenario 2) described in the main text. Here the TC distribution is well described by a t-Student distribution. The sample of TC has mean $= -0.71$, standard deviation $=0.57$, skewness $=-0.047$, and kurtosis $=6.2$. The t-Student fit has mean $=-0.71\pm 0.01$, standard deviation $=0.46\pm 0.01$, and degrees of freedom $=5.7\pm 0.6$. (c) Histogram and Gaussian fit corresponding to the scenario 3) described in the main text. Here the TC distribution is {\it not} well described by a normal distribution. The sample of TC has mean $= -0.40$, standard deviation $=1.1$, skewness $=0.88$, and kurtosis $=8.2$. The Gaussian fit has mean $=-0.40\pm 0.02$ and standard deviation $=1.1\pm 0.02$. (d) Histogram and t-Student fit corresponding to the scenario 4) described in the main text. Here the TC distribution is {\it not} well described by a t-Student distribution. The sample of TC has mean $= -0.39$, standard deviation $=1.6$, skewness $=1.1$, and kurtosis $=22$. The t-Student fit has mean $=-0.53\pm 0.02$, standard deviation $=0.85\pm 0.02$, and degrees of freedom $=2.4\pm 0.1$. Errors here correspond to the $95\%$ confidence intervals. Parameter values are $\gamma=1$, $\eta=1$, $\varepsilon=1$, $\sigma=1$, $\Delta t=1$, $p_0=1$, $\xi=1$, $\rho=1/2$. The strategy used has been the optimal trading strategy derived from the AC framework for $K=13$, $\lambda=0.3$.}
\label{TChistogram}
\end{center}
\end{figure}

In figure \ref{TChistogram} we show probability distributions of transaction costs under different price dynamics and impact models. Given these, the first questions we may ask are the following: If financial returns follow a particular distribution, what is the distribution followed by TC? Will it be similar to the one followed by returns? What is the effect of impact models on the TC distribution? Do they just alter the mean/variance, with respect to the distribution of returns, or do they also have an effect on the shape of the distribution (e.g. skewness)? We address these questions under different scenarios:
\begin{enumerate}[1)]
\item Price dynamics given by (\ref{pkAC}), with impact model determined by (\ref{ACPMI})-(\ref{ACTMI}) and $\chi \sim {\cal N}(0,1)$ (variable following a normal distribution with mean $0$ and standard deviation~$1$).
\item Price dynamics given by (\ref{pkAC}), with impact model determined by (\ref{ACPMI})-(\ref{ACTMI}) and $\chi \sim t_5(0,1)$ (variable following a t-Student\footnote{\label{footnoteStudent}The t-Student distribution, as we define it, has the following probability density function
\beq
f(x) = \frac{\Gamma\left( \frac{\nu+1}{2} \right)}{\sigma \sqrt{\nu\pi}\;\Gamma\left(\frac{\nu}{2}\right)} \left[ \frac{\nu+\left(\frac{x-\mu}{\sigma}\right)^2}{\nu} \right]^{-\left( \frac{\nu+1}{2} \right)}
\eeq
Where $\mu \equiv$ mean, $\sigma \sqrt{\nu/(\nu-2)} \equiv$ standard deviation, $\nu \equiv$ number of degrees of freedom, $\Gamma(x) \equiv$ Gamma function of $x$.
} 
distribution with 5 degrees of freedom, mean $0$ and standard deviation~$\sqrt{5/3}$).
\item Price dynamics given by (\ref{DMimpactmodel}), with impact model ${\cal I}(n_k)=-\xi \gamma n_k e^{-\rho \; k \Delta t}$, and $\zeta_k = \sigma \sqrt{\Delta t} \; \chi_k$, being $\chi \sim {\cal N}(0,1)$ (variable following a normal distribution with mean $0$ and standard deviation~$1$).
\item Price dynamics given by (\ref{DMimpactmodel}), with impact model ${\cal I}(n_k)=-\xi \gamma n_k e^{-\rho \; k \Delta t}$, and $\zeta_k = \sigma \sqrt{\Delta t} \; \chi_k$, being $\chi \sim t_5(0,1)$ (variable following a t-Student distribution with 5 degrees of freedom, mean $0$ and standard deviation~$\sqrt{5/3}$).
\end{enumerate}

Here $\xi = -1$ for a buy order, $\xi=1$ for a sell order, $\gamma$ and $\rho$ are constant parameters, and $k$, $\Delta t$, and $\sigma$, are as defined in the previous chapter (c.f. caption figure \ref{TChistogram} for parameter values used here). In all these cases, we perform a hypothesis test\footnote{Here we have used a Kolmogorov-Smirnov test. For the normality tests we have verified the hypothesis with Anderson-Darling, Jarque-Bera, and Lilliefors hypothesis tests.} with the null hypothesis that $c$ follows a normal distribution when $\chi\sim {\cal N}(0,1)$ and a t-Student distribution when $\chi\sim t_5(0,1)$. At a $1\%$ significance level, the hypothesis is not rejected in the first two cases, and rejected in the last two cases. This points to the fact that within the AC framework TC follow a similar distribution to financial returns. When returns follow a normal distribution with zero mean and standard deviation $\sigma$, the mean and variance of TC in the AC framework are given by
\beq
\mathbb{E}[c^{{}_{\rm AC}}] &=& - \frac{\xi}{N} \sum_{k=1}^K \left\{ x_k \Delta t \; g\left( \frac{n_k}{\Delta t} \right) + n_k h\left( \frac{n_k}{\Delta t}  \right) \right\} \label{ExAC} \\
\mathbb{V}[c^{{}_{\rm AC}}] &=& \frac{\sigma^2 \Delta t}{N^2} \sum_{k=1}^K x_k^2 \label{VxAC}
\eeq

In contrast, when considering the price evolution given by (\ref{DMimpactmodel}), TC follow a fundamentally different distribution to financial returns. These conclusions can be also derived through direct observation of equations (\ref{TCAC}) and (\ref{TCIMDM}): The first involves the sum of a set of variables $\chi$, while the second involves the sum of the product of these variables. Specially interesting is the fact that the price dynamics (\ref{DMimpactmodel}), together with the impact model ${\cal I}(n_k)=-\xi \gamma n_k e^{-\rho \; k \Delta t}$, seem to introduce a (positive) skewness in the distribution of TC, i.e. this model is able to describe asymmetric TC distributions, or in other words, it is flexible enough to reproduce the non-zero skewness regularly observed in practice.

\section{The utility function}

As we have seen, the TC distribution highly depends on the characteristics of market returns and market impact. Different markets have distinct properties, some feature distributions of returns with very fat tails, and in some the distributions are very asymmetric. Moreover, how our trading moves the security's price may be very different, depending on factors such as market liquidity. As such, not only is it important to examine various impact models, but also to investigate different utility functions. In particular, the optimal strategy obtained through the AC utility function (UF) fully depends on the mean and the variance of the TC distribution, only. As a result, markets with similar TC mean and variance, but with large asymmetries, or with frequent rare events, might give rise to very similar optimal execution strategies under the AC framework, but in reality, it might be advantageous (from the point of view of PnL) to execute differently in these markets. Consequently, it becomes natural to think about {\it ``transaction costs beyond mean-variance"}, i.e. to develop a formalism to account for properties of the TC distribution beyond its mean and its variance. One option would be to include subsequent moments (or rather cumulants) of the TC distribution in the UF. However, such ``cumulant expansion" is only a --better than AC-- approximation to account for the full distribution, characterised by the probability density function of TC, $f[c]$. In principle, we could consider this object, and obtain the execution strategy that ``optimises its shape". Following this rationale, we introduce the following utility function:
\newpage
\beq \label{DMUF}
U[c] = -(1-\lambda) \; \mathbb{E}[c] + \lambda \int_{-\infty}^{\tilde{c}} f[c] dc
\eeq
Here we have considered the integral of the TC distribution from $-\infty$ to $\tilde{c}$: a threshold below which we want to minimise TC. Optimising a UF including this integral (and depending on the value of $\tilde{c}$) will penalise more or less extreme negative costs\footnote{Notice that for $\tilde{c}>0$ one would be penalising (some) positive costs as well.}. This should be contrasted with the effect of the variance term in the AC utility function (\ref{ACutility}), which, in turn, penalises the width of the distribution, i.e. strategies in which TC {\it (positive or negative)} are very dissimilar. 

The first term (expectation of TC) in the UF (\ref{DMUF}) is still very relevant: it penalises the the speed of execution; if we were to minimise the utility function $U[c] = \int_{-\infty}^{\tilde{c}} f[c] dc$ (equivalent to $\lambda = 1$ in (\ref{DMUF})), we would obtain the optimal solution $n_1 = N$, $n_{k} = 0 \; \forall \; k\geq 2$, i.e. the execution of all shares in the first time interval following the order-start time. The {\it market impact} term $\sim \mathbb{E}[c]$ ``competes" with the integral term $ \sim \int_{-\infty}^{\tilde{c}} f[c] dc$, which captures the contribution of all moments (or cumulants) on the negative side of the TC distribution, and can thus be seen as an alternative {\it timing risk} measure. As we will see below, working with the UF (\ref{DMUF}) enables optimal strategies that execute the largest proportion of the shares in intervals later in time, a feature not possible under the AC formalism. Important is as well to consider the effect of the additional parameter $\tilde{c}$: depending on the value $\tilde{c}$ (the magnitude of the tail of the TC distribution considered in the UF) we will see a crossover from optimal strategies trading more shares in earlier time intervals to optimal strategies trading more shares later in time. Finally, it is noteworthy that (\ref{DMUF}) is just an example of a wider class of utility functions that can be considered to take into account the effect of the higher moments (or the full TC distribution) into the optimal strategy. More concretely, two cases of interest are
\beq
U[c] &=& -(1-\lambda) \; \mathbb{E}[c] + \frac{\lambda}{2} \int_{-\infty}^{-|\tilde{c}|} f[c] dc + \frac{\lambda}{2} \int_{|\tilde{c}|}^{\infty} f[c] dc \label{UFclass1} \\
U[c] &=& -(1-\lambda) \; \mathbb{E}[c] + \lambda \int_{-|\tilde{c}|}^{|\tilde{c}|} f[c] dc \label{UFclass2}
\eeq
In the first one, both positive and negative tails of the TC distribution are considered, while in the second, only the body of the TC distribution in taken into account. However, in-depth investigation of these instances is beyond the scope of this work, and it is left for for future research.

\newpage

\section{Optimisation method}

Given the utility function (\ref{DMUF}), we can now proceed to numerically obtain the optimal trading strategy that is derived from it. To start with, we consider the price evolution and impact model considered by Almgren and Chriss (equations (\ref{pkAC}) and (\ref{ACPMI})-(\ref{ACTMI}), with $\chi$ a Gaussian variable). In the probability density function $f[c]$ we have, in this case, the (stochastic) TC given by (\ref{TCAC}). In order to minimise the UF, we consider $10000$ realisations of the vector of random variables\footnote{Unless stated differently, this will be the case for all simulations presented throughout the text.} $(\chi_1,\ldots,\chi_K)$, which, for a given trading strategy $\{n_k\}$ (with ${k=1,\ldots,K}$) give rise to corresponding price vectors $(p_1,\ldots,p_K)$, and finally to $10000$ TC, $c^{{}_{\rm AC}}$, distributed as shown in figure \ref{TChistogram}(a). Provided the random samples of $c^{{}_{\rm AC}}$ have sufficiently converged to the underlying probability distribution\footnote{This is an important point that depends on the sample size and on the form of the underlying distribution, and which we will discuss further below.}, as we consider different realisations of $\{n_k\}$, the UF (\ref{DMUF}) will reach a minimum for one of them, which will be the {\it optimal trading strategy} (or {\it optimal execution strategy}), $\{n_k^{\rm opt}\}$. 

\begin{figure}[t]
\begin{center}
\includegraphics[width=\linewidth]{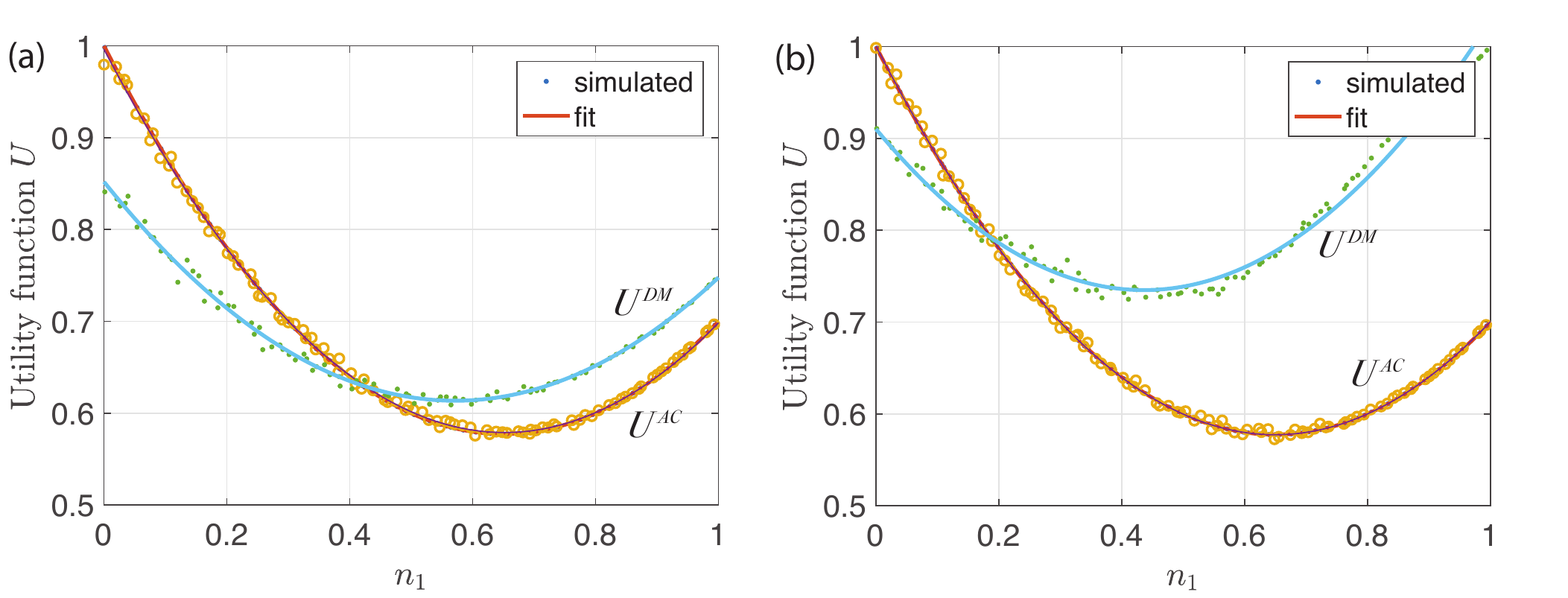}
\caption[Utility function for 2 trading intervals.]{Utility function in the case in which the trading horizon is subdivided into of 2 time intervals ($K=2$). (a) Case $\tilde{c}=-1$. Utility function (\ref{ACutility}) derived from the Almgren-Chriss formalism ($U^{{}_{\rm AC}}$). Here we have plotted both the ``analytical solution" (obtained via the equations (\ref{ExAC})-(\ref{VxAC})) and the ``numerical solution" (obtained via numerical simulation of the TC distribution), labelled $U^{{}_{\rm AC}}$. Both curves almost overlap. The UF derived from equation (\ref{DMUF}), labelled $U^{{}_{\rm DM}}$, is also shown. In this scenario, both curves show a minimum (optimal trading strategy) with $n_1 > 0.5$.
(b) Case $\tilde{c}=-0.5$. In this scenario, while $U^{{}_{\rm AC}}$ shows a minimum with $n_1 > 0.5$, this is not the case for $U^{{}_{\rm DM}}$, which shows a minimum with $n_1 < 0.5$. The ``analytical" and ``numerical" solution to the AC utility function almost overlap.
In both plots, the dots correspond to the values of the utility function computed over a quasi-random sample. The solid lines correspond to a fit of these points to a quadratic polynomial.
The Monte Carlo sample contains $100$ quasi-random strategies in each case, and $10000$ price paths have been used to estimate the UF at every point. The other parameter values are $\gamma=1$, $\eta=1$, $\varepsilon=1$, $\sigma=1$, $\Delta t=1$, $\xi=1$, $\lambda = 0.3$. The number of shares $n_k$, executed in time interval $k$ is expressed as a ratio to the order size, i.e. we take $N=1$.}
\label{UF2D}
\end{center}
\end{figure}

\begin{figure}[t]
\begin{center}
\includegraphics[width=\linewidth]{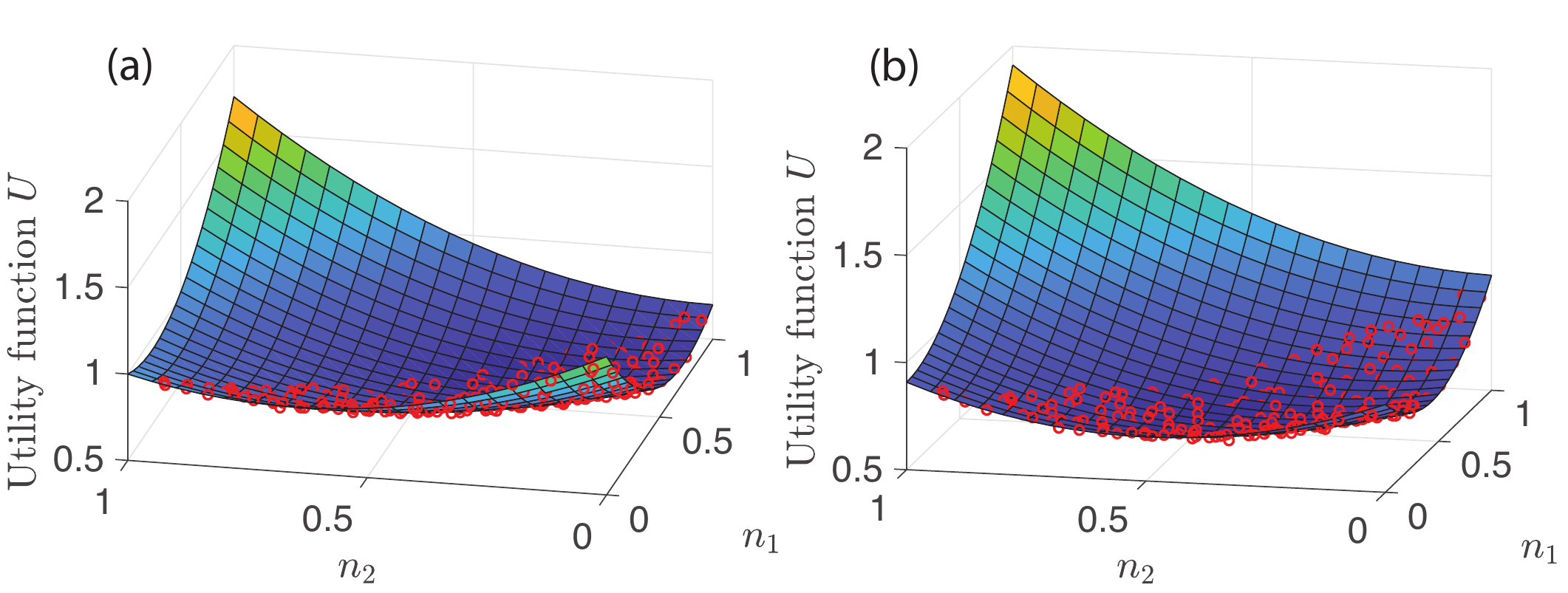}
\caption[Utility function for 3 trading intervals.]{Utility function in the case in which the trading horizon is subdivided into of 3 time intervals ($K=3$), and with $\tilde{c}=-0.5$. (a) Utility function (\ref{ACutility}) derived from the Almgren-Chriss formalism, and obtained via numerical simulation of the TC distribution. The result obtained via the analytical solution is visually almost indistinguishable. (b) Utility function derived from equation (\ref{DMUF}). In both plots, the dots correspond to the values of the UF computed over a quasi-random sample. The surfaces correspond to a fit of these points to a quadratic polynomial. Notice that, although shown here, the fitting curve fulfilling $n_1+n_2>N$ is actually not defined in the context of our problem: due to the constraint $\sum_{k=1}^K n_k = N$, any pair of two variables must fulfil $n_1+n_2 \leq N$.
The Monte Carlo sample contains $\sim 200$ quasi-random strategies in each case, and $10000$ price paths have been used to estimate the UF at every point. The other parameter values are $\gamma=1$, $\eta=1$, $\varepsilon=1$, $\sigma=1$, $\Delta t=1$, $\xi=1$, $\lambda = 0.3$. The number of shares $n_k$, executed in time interval $k$ is expressed as a ratio to the order size, i.e. we take $N=1$.}
\label{UF3D}
\end{center}
\end{figure}

In order to minimise the utility function, one may be inclined to use an algorithm of the type ``gradient descent" \cite{Cauchy47}. These, however, highly depend on the initial point from which we start the search of the global minimum, and might yield optimal solutions which correspond to local minima. In particular, due to the stochastic nature of our UF, and depending on the level of convergence of the considered sample of TC to the subjacent probability distribution, the (numerically-estimated) utility function is ``noisy", i.e. it might have a multitude of local minima in the multi-dimensional space span by the variables $\{n_1,\ldots,n_K\}$. As a matter of fact, in the results below we show how we have encountered this issue. In order to circumvent this difficulty, we can use a ``Monte Carlo optimisation" method: A sample of {\it quasi-random numbers}\footnote{Quasi-random numbers present a higher degree of correlation than a standard random sample, i.e. they fill the corresponding hypercube more uniformly.} $\{n_1,\ldots,n_K\}$, fulfilling the constraint $\sum_{k=1}^K n_k = N$, is generated inside the hypercube with vertices $\{v_1,\ldots,v_K\}$, with $v_j \in \{0,N\}$ ($j=1,\ldots,K$). We compute the value of the UF at each of these points, and select among these many the vector $\vec{n}$ giving rise to the lowest utility. Although, in our case, this procedure proves to be more effective than gradient-descent methods in order to find the global minimum, due to the noisy nature of the UF, the precision to which we determine the optimal solution may be inadequate. This issue is illustrated in figure \ref{UF2D}. However, we postulate that the UF should be a smooth (not noisy) function of the variables $\{n_1,\ldots,n_K\}$, and because our Monte Carlo simulations seem to indicate that the UF corresponding to the subjacent TC distribution does have only one local minimum, we fit the numerically-obtained values of the UF in the subspace span by the variables $\{n_1,\ldots,n_{K-1}\}$ to a $(K-1)$-dimensional {\it quadratic polynomial}. In figures \ref{UF2D} and \ref{UF3D} we see the fitted curve for the cases in which the trading horizon is subdivided into two and three time intervals, respectively. Once we have this curve, we can use an optimisation method to find the global minimum: yet again a quasi-random minimisation or a gradient-descent procedure, both of which work well for this smooth utility function. As we show in tables \ref{TabOptSols2D} and \ref{TabOptSols3D}, the best accuracy proves to be obtained with the latter method, and will we take this solution as the optimal execution strategy.

\begin{table}[t]
\begin{center}
\resizebox{\columnwidth}{!}{
\begin{tabular}{|c|c|c||c|c||c|c||c|c|}
\cmidrule[1pt]{2-9}                  
\multicolumn{1}{c}{}&\multicolumn{2}{|c||}{{\bf AC analytic}}&\multicolumn{2}{|c||}{{\bf AC numeric}}&\multicolumn{2}{|c||}{\bf{DM $\bm{(\tilde{c}=-1)$}}} &\multicolumn{2}{|c|}{\bf{DM $\bm{(\tilde{c}=-0.5)$}}} \\ 
\midrule
{\bf Opt. method} & $\bm{\vec{n}^{\rm opt}}$ & $\bm{U^{\rm opt}}$ & $\bm{\vec{n}^{\rm opt}}$ & $\bm{U^{\rm opt}}$ & $\bm{\vec{n}^{\rm opt}}$ & $\bm{U^{\rm opt}}$ & $\bm{\vec{n}^{\rm opt}}$ & $\bm{U^{\rm opt}}$ \\\midrule
GD & $(0.65,0.35)$ & $0.58$ & $(0.56,0.44)$ & $0.58$ & $(0.56,0.44)$ & $0.60$ & $(0.71,0.29)$ & $0.81$ \\
MC & $(0.65,0.35)$ & $0.58$ & $(0.61,0.39)$ & $0.57$ & $(0.56,0.44)$ & $0.61$ & $(0.41,0.59)$ & $0.72$ \\
Q2 Fit + MC & $(0.65,0.35)$ & $0.58$ & $(0.65,0.35)$ & $0.58$ & $(0.57,0.43)$ & $0.61$ & $(0.44,0.56)$ & $0.73$ \\
Q2 Fit + GD & $(0.65,0.35)$ & $0.58$ & $(0.65,0.35)$ & $0.58$ & $(0.57,0.43)$ & $0.61$ & $(0.44,0.56)$ & $0.73$ \\\midrule
\end{tabular}
}
\caption[Optimal trading strategies for 2 trading intervals.]{Solutions, expressed in the form $(n_1,n_2)$, and value of the UF, for the optimal execution strategy obtained by various optimisation methods, in the case in which the trading horizon is subdivided into of 2 time intervals ($K=2$). GD refers to `Gradient-descent' (the solution obtained by applying standard optimisation methods \cite{Cauchy47} to the UF), MC refers to `Monte Carlo' (the solution obtained evaluating the UF at a quasi-random sample of strategies), and Q2 Fit refers to `Quadratic Fit' (the solution obtained from the methods MC and GD applied to the UF approximated by a quadratic-polynomial fit to the quasi-random sample). AC denotes the solution obtained by using the AC utility function (\ref{ACutility}), which can be expressed analytically, or solved by numerical simulation of the TC distribution. DM denotes the solution obtained by using the UF (\ref{DMUF}). The Monte Carlo sample contains $100$ quasi-random strategies in each case. We have investigated the cases corresponding to $\tilde{c} = -1$ and $\tilde{c} =-0.5$. The other parameter values are $\gamma=1$, $\eta=1$, $\varepsilon=1$, $\sigma=1$, $\Delta t=1$, $\xi=1$, $\lambda = 0.3$. The number of shares $n_k$, executed in time interval $k$, is expressed as a ratio to the order size, i.e. we take $N=1$.}
\label{TabOptSols2D}
\end{center}
\end{table}

~

Figure \ref{UF2D} presents various important insights. First, the UF (\ref{DMUF}) does indeed present a local minimum. This means that our modelling of timing risk is meaningful in the context of the formulated optimisation problem, in the sense that it competes with market impact, and it is in the balance between the two that we find an optimal execution strategy. Second, by comparing figure \ref{UF2D}(a) and \ref{UF2D}(b), we see that, while the AC utility function always gives rise to an optimal trading strategy with $n_1>0.5$ in the case of 2 trading intervals, the UF (\ref{DMUF}) may present solutions with both cases, $n_1>0.5$ or $n_1<0.5$, depending on the value of $\tilde{c}$. Indeed, the parameter $\tilde{c}$ plays an essential role: as we increase the value of $\tilde{c}$, we ``redefine" timing risk to include --not only the negative tail of the TC distribution (rare losses)-- but also the body of the distribution. This redefinition of timing risk from ``extreme loss" to ``loss" effectively penalises less being patient to complete the order, and importantly it admits solutions in which we trade a larger proportion of the order in later time intervals. Consequently, we find a crossover from $n_1>0.5$ to $n_1<0.5$ when the value of $\tilde{c}$ is increased from $-1$ to $-0.5$. Finally, it is important to remark that the stochastic nature of the problem leads to a noisy simulated utility function (c.f. dots in figure \ref{UF2D}). In this respect, in order to get an accurate solution to the problem, it is important to simulate enough price paths so that the simulated TC distribution has sufficiently converged to the underlying probability distribution. In the case simulated here, judging from the values of the moments of the distribution --as compared to the theoretical values-- one can observe a large degree of convergence: the theoretical values of the first four moments of the distribution are, in this case, mean $=0.71$, standard deviation $=0.44$, skewness $=0$, kurtosis $=3$, which are in close agreement with the values obtained from our simulation (see caption figure \ref{TChistogram}). Although this convergence is certainly important, the introduced procedure of determining the optimal trading strategy via a fitted curve to the simulated UF makes the obtained solution robust against statistical fluctuations, i.e. a second-order polynomial fit to even more noisy utility functions proves to deliver similar optimal execution strategies. This will allow us to tackle more complicated scenarios in which the mentioned convergence is not as accurate, such as the cases (b), (c) and (d) shown in figure \ref{TChistogram}. Furthermore, obtaining the optimal trading strategy from the fitted UF also allows us to reduce the density of simulated points: while obtaining the minimum from the individual evaluation of the utility function at each of the simulated points requires a high density of candidate strategies to accurately determine the local minimum, rather similar polynomial fits are obtained when the density of points is decreased, and so the obtained minimum is relative robust in this sense as well.

~

In figure \ref{UF3D} we show the dependency of the UF on the number of shares executed in the first and second time interval, in the case in which the trading horizon is subdivided into three time intervals $(K=3)$. Similarly to the case $K=2$ discussed above, both utility functions, (\ref{ACutility}) and (\ref{DMUF}), have a local minimum. In figure \ref{UF3D}(a) we show the simulation of $~200$ strategies (red dots) following the AC price dynamics, and the corresponding quadratic fitting curve. In figure \ref{UF2D}(b) we show the same for the UF (\ref{DMUF}). Visually, both fitting approximations are rather accurate, and thus we yet again use them to determine the local minimum. Doing so we obtain the results presented in table \ref{TabOptSols2D}. The best approximation to the optimal trading strategy is provided by the optimisation Q2 Fit + GD (quadratic fit followed by gradient-descent). As in the case for two trading intervals, we find a ``reversal" in the optimal execution as we vary the parameter $\tilde{c}$, from trading the largest proportion of the shares in the earliest time intervals, to trading the largest proportion of the shares in the latest time intervals. Finally, it is important to remark that when generating quasi-random points corresponding to different strategies, we must do so by respecting the constraint $\sum_{k=1}^K n_k = N$. Consequently, we have written $n_K = N - \sum_{k=1}^{K-1} n_k$, and taken a sample of vectors $(n_1,\ldots,n_{k-1})$ such that
\beq \label{Kminus1constraint}
\sum_{k=1}^{K-1} n_k \leq K
\eeq
This is why in figure \ref{UF3D} we observe that all the randomly generated points rest within the triangular region corresponding to $n_1+n_2 \leq N$. Importantly, such constraint has significant implications for the scaling of the problem towards a higher dimensionality (number of time intervals in which the trading horizon is subdivided). In order to generate $q$ quasi-random numbers in the region fulfilling (\ref{Kminus1constraint}), we must first generate $Q>q$ quasi-random numbers in the hypercube with vertices $\{v_1,\ldots,v_{K-1}\}$, with $v_j \in \{0,N\}$ ($j=1,\ldots,K-1)$ and then post-select for the fitting the subsample that satisfies (\ref{Kminus1constraint}). This implies that, if we want to preserve an approximately constant number of points used for the fitting as the dimensionality of the problem is increased, the number $Q$ of pre-generated random strategies needs to increase exponentially with $K$. We numerically observe that for $K\geq 12$, $Q$ needs to be larger than $10^8$. On top of this, we expect that, as $K$ increases, the goodness-of-fit of the polynomial curve will worsen if we keep $q$ constant\footnote{How does $q$ need to be increased as we increase the dimensionality --if we would like the fitting curve to remain accurate-- can be investigated e.g. by monitoring the p value of a goodness-of-fit test as we increase the dimensionality of the problem.}. Therefore, $q$ should also scale with the dimensionality $K$. This leads to a limitation of our approach for increased granularity of the trading intervals, and in practice when $K\gtrsim5$. How to improve the scaling of this problem is left as a topic for future research.  

\begin{table}[t]
\begin{center}
\resizebox{\columnwidth}{!}{
\begin{tabular}{|c|c|c|c|c|}
\cmidrule[1pt]{1-5}                  
{\bf Opt. method} & $\bm{\vec{n}^{\rm opt}}$ {\bf AC analytic} & $\bm{\vec{n}^{\rm opt}}$ {\bf AC numeric}& $\bm{\vec{n}^{\rm opt}}$ \bf{DM $\bm{(\tilde{c}=-1)$}} & $\bm{\vec{n}^{\rm opt}}$ \bf{DM $\bm{(\tilde{c}=-0.5)$}} \\ 
\midrule
GD & $(0.60,0.26,0.14)$ & $(0.79,0.17,0.04)$ & $(0.35,0.32,0.33)$ & $(0.79,0.17,0.04)$ \\
MC & $(0.59,0.27,0.14)$ & $(0.59,0.27,0.14)$ & $(0.36,0.31,0.33)$ & $(0.36,0.34,0.30)$ \\
Q2 Fit + MC & $(0.59,0.27,0.14)$ & $(0.59,0.27,0.14)$ & $(0.43,0.32,0.25)$ & $(0.27,0.35,0.38)$ \\
Q2 Fit + GD & $(0.60,0.26,0.14)$ & $(0.60,0.26,0.14)$ & $(0.40,0.32,0.28)$ & $(0.29,0.34,0.37)$ \\
\midrule
\end{tabular}
}
\caption[Optimal trading strategies for 3 trading intervals.]{Solutions, expressed in the form $(n_1,n_2,n_3)$, for the optimal execution strategy obtained by various optimisation methods, in the case in which the trading horizon is subdivided into of 3 time intervals ($K=3$). GD refers to `Gradient-descent' (the solution obtained by applying standard optimisation methods \cite{Cauchy47} to the UF), MC refers to `Monte Carlo' (the solution obtained evaluating the UF at a quasi-random sample of strategies), and Q2 Fit refers to `Quadratic Fit' (the solution obtained from the methods MC and GD applied to the UF approximated by a quadratic-polynomial fit to the quasi-random sample). AC denotes the solution obtained by using the AC utility function (\ref{ACutility}), which can be expressed analytically, or solved by numerical simulation of the TC distribution. DM denotes the solution obtained by using the UF (\ref{DMUF}). The Monte Carlo sample contains $\sim 200$ quasi-random strategies in each case. We have investigated the cases corresponding to $\tilde{c} = -1$ and $\tilde{c} =-0.5$. The other parameter values are $\gamma=1$, $\eta=1$, $\varepsilon=1$, $\sigma=1$, $\Delta t=1$, $\xi=1$, $\lambda = 0.3$. The number of shares $n_k$, executed in time interval $k$, is expressed as a ratio to the order size, i.e. we take $N=1$.}
\label{TabOptSols3D}
\end{center}
\end{table}

\section{Optimal trading strategy}

We next examine the solution for the optimal trading strategy as we increase the number of trading intervals and vary the parameters of the utility function. As we have mentioned above, using our method in high dimensions (large number of time intervals that subdivide the trading horizon) should be handled with care, verifying:
\begin{enumerate}[i)]
\item That the distribution of generated transaction costs has sufficiently converged to the underlying probability distribution, to avoid too much noise in the utility function.
\item That the sample of generated strategies is sufficiently dense in the $(K-1)$-hypercube, to assure an accurate representation of the utility function.
\item That the polynomial fit to the strategy points passes an appropriate goodness-of-fit test.
\end{enumerate}
For $K=2$ and $K=3$, following the discussion above and the illustration of figures \ref{UF2D} and \ref{UF3D}, it is easy to see that these conditions are fulfilled, and --due to the high precision of the results in these cases-- to presume that we will have a sufficient degree of accuracy in slightly higher dimensions. However, in full rigour the conditions above should be mathematically verified, in particular when $K \gtrsim 5$.
Furthermore, it is important to notice that how well these conditions are fulfilled will depend --not only on the utility function-- but also on the distribution of financial returns, as well as on the impact model assumed in the price evolution. In particular, in figure \ref{TChistogram} we see that considering t-Student returns affects only slightly the convergence of the TC sample to the underlying probability distribution: the theoretical values of the first four moments of the distribution associated to panel b) are: mean $=0.71$, standard deviation $=0.57$, skewness $=0$, kurtosis $=6$ (c.f. caption of figure \ref{TChistogram} for the obtained values through the sample). Considering a different impact model, however, has a stronger impact on the convergence of the TC sample to the underlying probability distribution: generating various samples of the same size, we see that the obtained mean and standard deviation associated to panels c) and d) are somewhat robust, but not so robust are the skewness and kurtosis of the sample. This suggests that we should consider taking a larger sample size in the simulations of TC associated to the impact model used to derive these results. However, once we have made this this remark, one should notice that the convergence of the TC sample to the underlying probability distribution will be reflected in the noisy behaviour of the generated utility function, but --to our advantage-- the fitting function is somewhat robust with respect to the noise and density of generated trading strategies.

\newpage

\begin{figure}[t]
\begin{center}
\includegraphics[width=\linewidth]{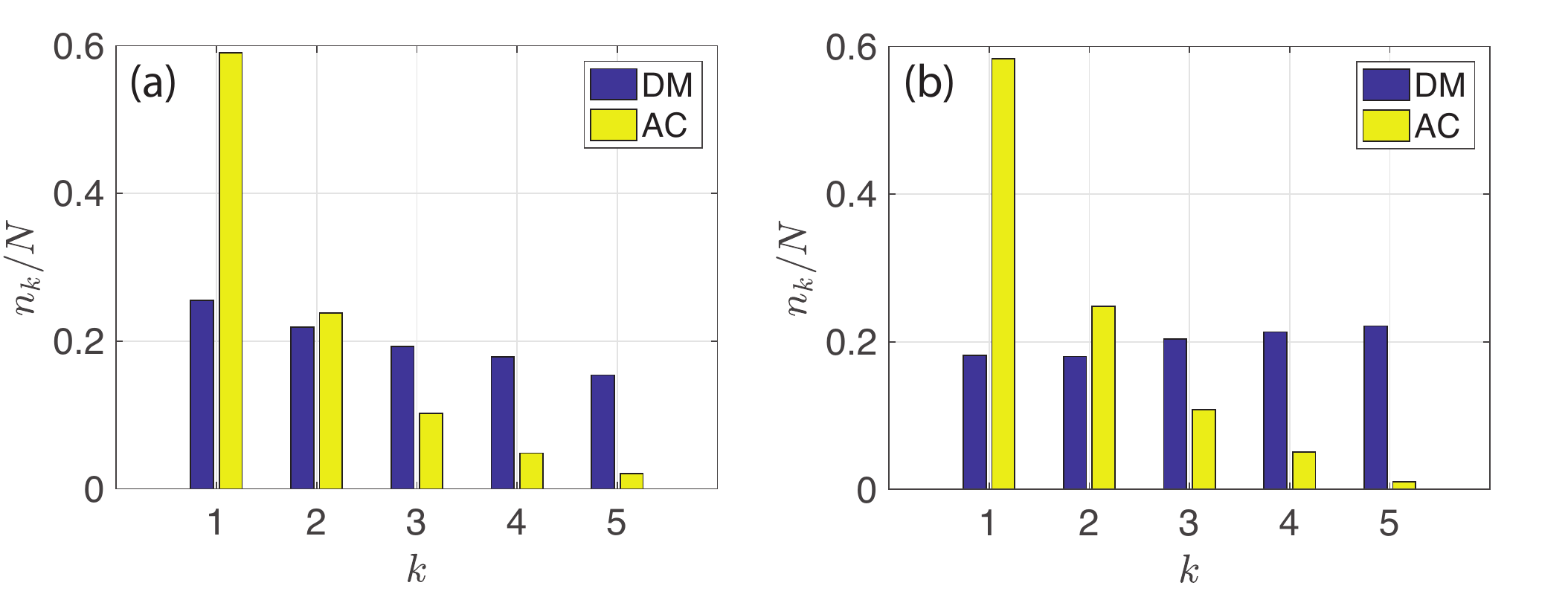}
\caption[Optimal trading strategy for 5 trading intervals.]{Number of shares $n_k$ sold in interval $k$, as a proportion to the total number of order shares $N$. Here 5 time intervals have been considered. (a) Optimal strategy derived via the UF (\ref{DMUF}) (DM) and via the UF (\ref{ACutility}) (AC), in the case $\lambda = 0.3$, $\tilde{c} = -1$. While the AC solution decays exponentially, the DM solution seems to decay linearly. (b) Optimal strategy derived via the UF (\ref{DMUF}) (DM) and via the UF (\ref{ACutility}) (AC), in the case $\lambda = 0.3$, $\tilde{c} = -0.5$. In this case, the strategy obtained via the UF (\ref{DMUF}) is almost constant but increases slightly the proportion of executed shares with the trading interval. The analytical AC solution (obtained using equations (\ref{ExAC})-(\ref{VxAC})) coincides with the numerical AC solution shown here (obtained via Monte Carlo simulation) to a very high --but finite-- precision (due to statistical fluctuations). The other parameter values are $\gamma=1$, $\eta=1$, $\varepsilon=1$, $\sigma=1$, $\Delta t=1$, $\xi=1$.}
\label{nkFig}
\end{center}
\end{figure}

Following the logic explained above, we here present results for the optimal trading strategy obtained with our utility function (\ref{DMUF}), using the AC price-impact model. In figure \ref{nkFig} we show the comparison between the optimal trading strategies obtained via equations (\ref{DMUF}) and (\ref{ACutility}). For the given values of $\tilde{c}$, the strategy derived by AC decays more rapidly with the trading interval than that obtained with the UF (\ref{DMUF}), i.e. under the AC model the trader's patience ``decays exponentially". However, in a framework in which timing risk is modelled through the negative tail of the TC distribution, the patience of the trader seems to decay more slowly (it appears to be {\it linear} with the time interval). More interestingly, at some level of $\tilde{c}$ --as we increase its value-- the number of traded shares no longer decreases with the trading interval. We conjecture that this is because at some level of $\tilde{c}$, the integral in the timing-risk term of equation (\ref{DMUF}) plays a different role (as the integral between $-\infty$ and $\tilde{c}$ of the probability density function captures --in that case-- almost the whole distribution). When the tail includes a larger proportion of the distribution, trading itself is penalised (as it incurs a negative transaction cost), thus a slight preference to leave a larger proportion of the shares to be executed in later intervals (case in which market impact has a contribution over less intervals, due to the decaying nature of the temporary market impact).

\newpage

In order to explore further this behaviour, we have calculated the optimal trading strategy across different values of $\lambda$ and $\tilde{c}$, computed with the utility function (\ref{DMUF}), here in the case of two trading intervals. The results are shown in table \ref{mapDMUDsols}. Starting with a fixed value of $\tilde{c}=-1$, we see that --as we increase the value of $\lambda$-- we go from a TWAP strategy, that distributes equally the number of executed shares among all trading intervals, to a strategy where we trade most of the order shares in the first time interval. This outcome is similar to what it is obtained with the UF (\ref{ACutility}), and intuitive: $\lambda\to 0$ implies that the market-impact term $\sim \mathbb{E}[c]$ dominates the UF, and this will be minimised in a situation where we provide the minimal ``shock" to the market per trading interval, namely when we trade at a constant rate through the time horizon. $\lambda\to 1$, in contrast, implies that the timing-risk term $\sim \int_{-\infty}^{\tilde{c}} f[c] dc$ dominates the UF. In this case, we aim to minimise the area under the tail of the TC distribution (extreme negative costs), which is accomplished by executing the order ``rapidly", i.e. completing it within the first time interval(s). More striking is the behaviour when $\tilde{c}$ is varied. As we increase $\tilde{c}$, for example going from $\tilde{c}=-1$ to $\tilde{c} = -0.5$, we yet again find a homogeneous distribution of executed shares among trading intervals for $\lambda\to 0$. However, for larger $\lambda$ we observe an ``inversion effect": Now most of the shares are executed in the latest time interval(s), i.e. the UF (\ref{DMUF}) incorporates an extra degree of ``patience" with respect to the AC utility function. This phenomenon is mostly relevant in the context of describing markets where most of the trading is completed towards the end of the day (due to increased liquidities at the close). In future research, it will be interesting to investigate the necessary degrees of freedom to be included in the model in order to recover the common ``U shape" (meaning that most of the trading is done at the beginning and at the end of the day) with lower trading volumes in intermediate time intervals\footnote{A potential option to accomplish this might be adding a variance term $\sim \mathbb{V}[c]$ to the UF (\ref{DMUF}), but exploring this avenue is left as a future prospect.}

\begin{table}[t]
\begin{center}
\begin{tabular}{c|c|c|c|c|c|}
\cline{2-6}
& $\bm{\tilde{c}=-1}$ & $\bm{\tilde{c}=-0.5}$ & $\bm{\tilde{c}=0}$ & $\bm{\tilde{c}=0.5}$ & $\bm{\tilde{c}=1}$ \\ \cline{1-6}
\multicolumn{1}{ |l  }{\multirow{1}{*}{$\bm{\lambda=1}$} } &
\multicolumn{1}{ |c| }{(1.00,0.00)} & (0.21,0.78) & (0.00,1.00) & (0.00,1.00) & (0.00,1.00)     \\ \cline{1-6}
\multicolumn{1}{ |l  }{\multirow{1}{*}{$\bm{\lambda=0.7}$} } &
\multicolumn{1}{ |c| }{(0.84,0.16)} & (0.33,0.67) & (0.21,0.79) & (0.38,0.62) & (0.47,0.53)     \\ \cline{1-6}
\multicolumn{1}{ |l  }{\multirow{1}{*}{$\bm{\lambda=0.5}$} } &
\multicolumn{1}{ |c| }{(0.66,0.34)} & (0.38,0.62) & (0.38,0.62) & (0.46,0.54) & (0.49,0.51)     \\ \cline{1-6}
\multicolumn{1}{ |l  }{\multirow{1}{*}{$\bm{\lambda=0.3}$} } &
\multicolumn{1}{ |c| }{(0.57,0.43)} & (0.44,0.56) & (0.45,0.55) & (0.48,0.52) & (0.49,0.51)     \\ \cline{1-6}
\multicolumn{1}{ |l  }{\multirow{1}{*}{$\bm{\lambda=0}$} } &
\multicolumn{1}{ |c| }{(0.50,0.50)} & (0.50,0.50) & (0.50,0.50) & (0.50,0.50) & (0.50,0.50)     \\ \cline{1-6}
\end{tabular}
\caption[Map of optimal trading strategies for 2 trading intervals.]{Solutions, expressed in the form $(n_1,n_2)$, for the optimal execution strategy obtained by minimising the UF (\ref{DMUF}). As with the AC utility function (for a given value of $\tilde{c}$) decreasing $\lambda$ gives a more ``homogeneous" solution ($n_1\approx n_2$). Actually, solutions for different values of $\lambda$ corresponding to $\tilde{c} \approx -1$ are somewhat similar to those obtained by AC. However, for a given value of $\lambda$, increasing $\tilde{c}$ changes the notion of the timing-risk term in the utility function. From penalising ``patience" at $\tilde{c} = -1$, it penalises ``impatience" for $\tilde{c} \gtrsim -0.5$. We thus find an ``inversion effect", from $n_1>n_2$ to $n_2>n_1$, when increasing $\tilde{c}$. The other parameter values are $\gamma=1$, $\eta=1$, $\varepsilon=1$, $\sigma=1$, $\Delta t=1$, $\xi=1$.}
\label{mapDMUDsols}
\end{center}
\end{table}

\begin{figure}[t]
\begin{center}
\includegraphics[width=0.95\linewidth]{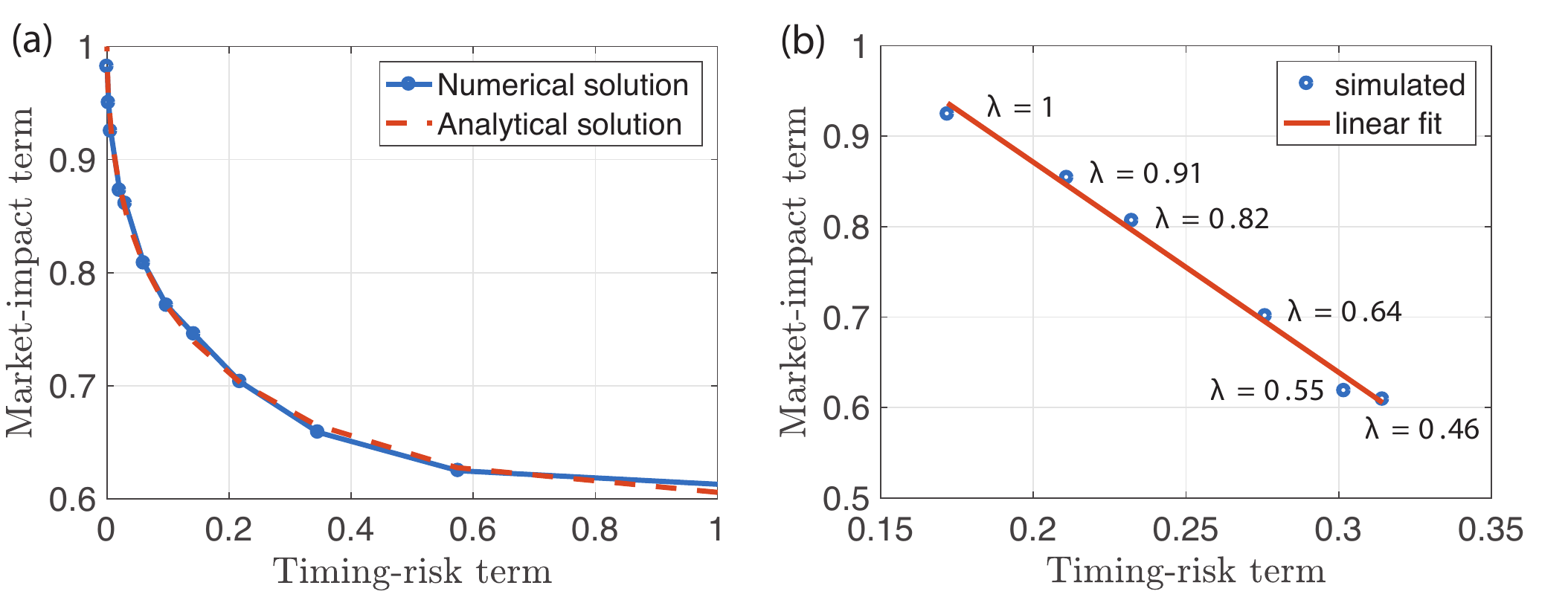}
\caption[Efficient trading frontier.]{Efficient trading frontier computed with 5 trading intervals. (a) Result of Almgren and Chriss, where the market impact -- timing risk relationship is nonlinear. In particular --for small values TC volatility-- taking twice as much timing risk might reduce market impact by significantly more than a factor of two (when both rescaled to the corresponding units). Here `Market-impact term' = -$\mathbb{E}[c]$, while `Timing-risk term' = $\mathbb{V}[c]$. (b) Result obtained with the UF (\ref{DMUF}) in the case $\tilde{c} = -1$. Within this range of values of $\lambda$, and for sufficiently small $\tilde{c}$, the relationship market impact -- timing risk is approximately linear. Here `Market-impact term' = -$\mathbb{E}[c]$, while `Timing-risk term' = $\int_{-\infty}^{\tilde{c}} f[c] dc$. Notice that the units of the Timing-risk term are different from each other in panels (a) and (b), and that these units are themselves different to the those of the Market-impact term. The other parameter values are $\gamma=1$, $\eta=1$, $\varepsilon=1$, $\sigma=1$, $\Delta t=1$, $\xi=1$.}
\label{EfficientFrontier}
\end{center}
\end{figure}

From the results above, we see that --at least for small $\tilde{c}$-- the timing-risk term $\sim \int_{-\infty}^{\tilde{c}} f[c] dc$ can be interpreted as a ``standard" TC risk (similar to TC volatility, or rather to TC value-at-risk). As such, we may wonder whether in a picture market impact vs. timing risk we can define an {\it `efficient trading frontier'}, similarly to how this is done within the AC formalism \cite{AlmgrenChriss01}, i.e. as the set of optimal trading strategies for different values of $\lambda$ in a market impact -- timing risk representation. In figure \ref{EfficientFrontier}(a) we display the result obtained for such efficient frontier when we use the AC utility function (\ref{ACutility}). Here we have computed the optimal execution strategy across different values of $\lambda \in [0,1]$ and plotted both, the solutions obtained via a numerical simulation of transaction costs, and using the analytical formulas (\ref{ExAC})-(\ref{VxAC}) for their expectation and variance. In figure \ref{EfficientFrontier}(b) we show the equivalent efficient frontier if we use the utility function (\ref{DMUF}). We do so for $\tilde{c} = -1$ and sufficiently large values of $\lambda$, when the integral term $\int_{-\infty}^{\tilde{c}} f[c] dc$ can be clearly interpreted as a ``standard" TC risk. Interestingly, in this case the efficient frontier displays a linear relationship between market impact and timing risk. In other words, if we are in a regime in which TC risk corresponds to extreme negative events, and if we scale MI and timing risk to the appropriate units, in order to reduce market impact by a factor of two, we need to take twice as much timing risk. Conversely, if we want to reduce timing risk by half, our execution will need to incur twice as much market impact. This is perhaps {\it the most} distinctive characteristic of timing risk modelled as done in the utility function (\ref{DMUF}), compared to a mean-variance UF of the type (\ref{ACutility}).

\section{Non-Gaussian returns}

One important assumption in the model of Almgren and Chriss is that, in their framework, returns are independent normally-distributed variables\footnote{More precisely, AC consider that {\it arithmetic} returns follow a Gaussian distribution.}. This enables, specifically, to obtain analytical formulas for TC expectation and variance, thus avoiding the need to numerically simulate the price dynamics (as done above via Monte Carlo). However, real markets are not that gentle, and it is common that the probability distribution of financial returns is leptokurtic (i.e. with fat tails) \cite{Bouchaud09}. In particular, we have analysed over $200$ markets, with the purpose of establishing certain universality for the distribution of financial returns. By downloading daily prices over a period of around $30$ years, one can observe that generically (both arithmetic and geometric) returns fail tests of normality at the most common significance levels. More specifically, less liquid markets (e.g. emerging markets) tend to show more rare events or certain autocorrelation, giving rise to distributions with heavier tails than those of the normal distribution. The difficulty or establishing a ``universal law" that describes\footnote{In the sense that it {\it estimates} or {\it predicts}.} market returns lies in the fact that --apart from providing a good fit-- the conjectured distribution should: i) have the minimum number of fitting parameters (to avoid overfitting) and ii) apply over the broad class of markets that it aims to describe with the same parameter values (within a margin of error). Following our analysis, the distributions that, to our knowledge, accomplish this best are: {\it t-Student} distributions, piece-wise distributions (e.g. Gaussian body and Pareto or exponential tails), or of the class of {\it `alpha-stable'} distributions, such as the {\it Levy} distribution, or the {\it truncated Levy} distribution \cite{Bouchaud09}. We will, therefore, for the illustration of our method under more realistic market conditions, consider the case of market returns distributed according to a t-Student distribution\footref{footnoteStudent}.

~

As we have seen, the rationale of introducing the utility function (\ref{DMUF}), as compared to the UF (\ref{ACutility}) is to precisely capture TC timing risk as originated by extreme negative events that may occur during the period of ``not executing" the order. It is therefore intuitive that, in markets characterised by fat-tailed distributions, our UF plays a relevant role in determining how to optimally execute a market order. In particular, as we will show, this is perhaps {\it the most relevant} application of our theory: Given two orders following different TC distributions, one Gaussian and one fat-tailed (e.g. t-Student), with the same mean and variance, but different higher moments (as it happens with heavy-tailed distributions), should these two orders be executed equally over the trading horizon? It seems intuitive that, if the ``tail-risk" (probability of negative rare events) is different, the execution strategy for both orders should be, in some sense, different. The UF provided by Almgren and Chris will provide, however, the same solution for the optimal execution strategy (as it is apparent from the fact that only TC mean and variance enter this optimisation problem). In contrast, we show next how the UF (\ref{DMUF}) distinguishes between these two situations, and suggests that the second order should be traded more or less aggressively than the first, depending on the value of $\tilde{c}$, which determines the characterisation of timing risk as more or less ``rare events".

\begin{figure}[t]
\begin{center}
\includegraphics[width=0.65\linewidth]{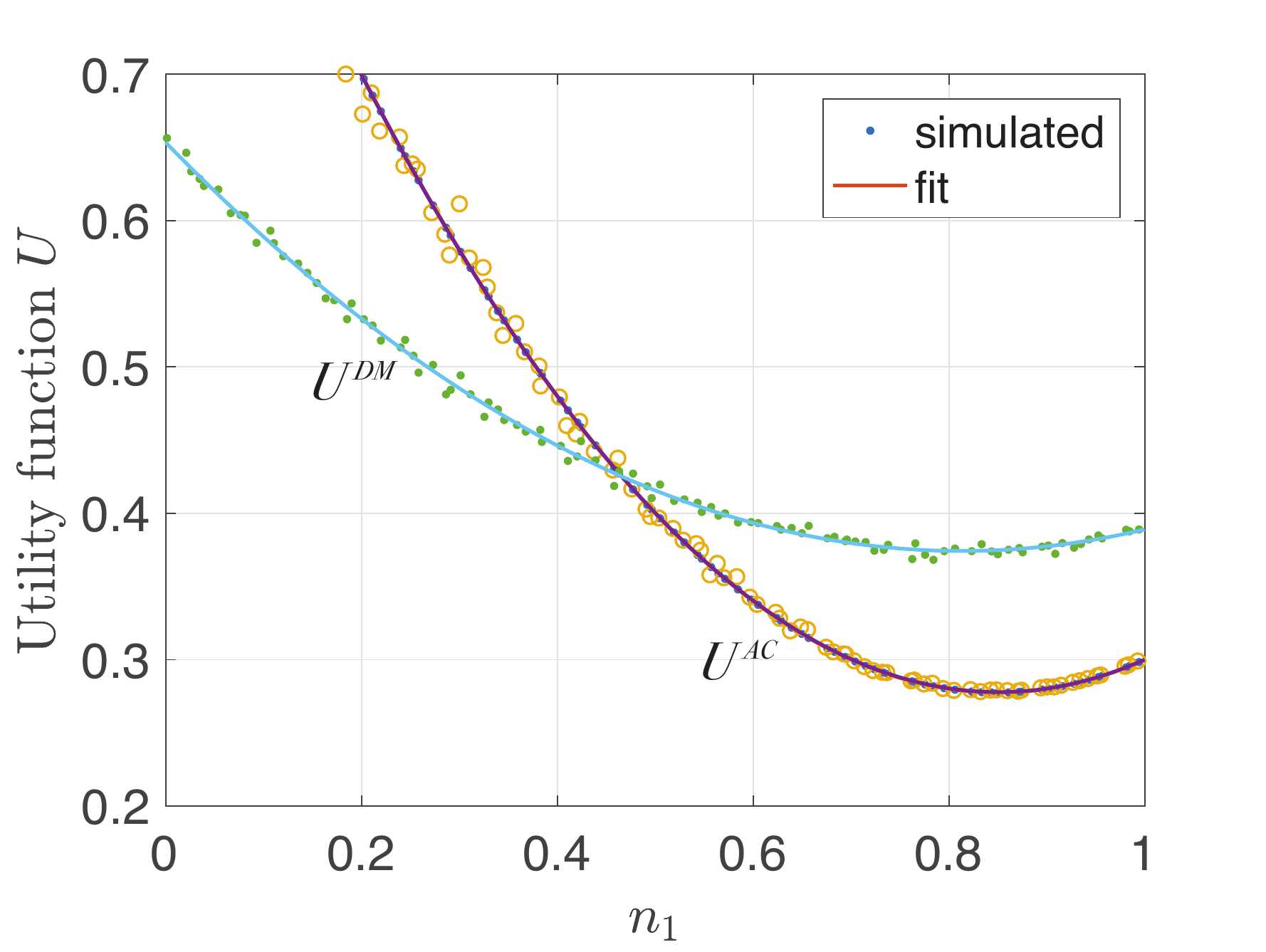}
\caption[Utility function for 2 trading intervals using t-Student returns.]{Utility function in the case in which the trading horizon is subdivided into 2 time intervals ($K=2$) and when the returns follow a t-Student distribution\footref{footnoteStudent} with $\mu=0$, $\sigma=\sqrt{(\nu-2)/\nu}$, and $\nu=5$. Here we show the utility function (\ref{ACutility}) numerically estimated using t-Student returns (labeled $U^{{}_{\rm AC}}$), and compared to the equivalent analytical result (which corresponds to Gaussian returns with $\sigma = 1$). Both curves overlap almost exactly. The utility function (\ref{DMUF}) estimated with t-Student returns ($\mu=0$, $\sigma=\sqrt{(\nu-2)/\nu}$, and $\nu=5$) is also shown (labeled $U^{{}_{\rm DM}}$).
The solutions for the optimal execution strategies (minima of the {\it fitted} UF), expressed in the form $(n_1,n_2)$ are $(0.85,0.15)$, $(0.85,0.15)$, and $(0.81,0.19)$, for AC analytical (Gaussian returns), AC numerical (t-Student returns), and corresponding to equation (\ref{DMUF}) (t-Student returns), respectively.
The dots correspond to the values of the utility function computed over a quasi-random sample. The solid lines correspond to a fit of these points to a quadratic polynomial.
The Monte Carlo sample contains $100$ quasi-random strategies in each case, and $10000$ price paths have been used to estimate the UF at every point. 
Parameter values are $\gamma=1$, $\eta=1$, $\varepsilon=1$, $\sigma=1$, $\Delta t=1$, $\xi=1$, $\lambda = 0.7$, $\tilde{c} = -1$. The number of shares $n_k$, executed in time interval $k$ is expressed as a ratio to the order size, i.e. we take $N=1$.}
\label{Student2D}
\end{center}
\end{figure}

In figure \ref{Student2D} we show the result of the simulation of $100$ strategies, for each of which the utility function has been evaluated using $10000$ price paths, with price dynamics given by (\ref{pkAC}), (\ref{ACPMI}), (\ref{ACTMI}), being now $\chi$ a t-Student-distributed variable\footref{footnoteStudent} with mean $0$, standard deviation $\sqrt{(\nu-2)/\nu}$, with $\nu=5$ degrees of freedom. We have taken this specific value of the $\sigma$ in order to make the standard deviation of the distribution (given by $\sigma \sqrt{\nu/(\nu-2)}$) equal to $1$, and thus compare the result with the one obtained using normally-distributed returns with mean $0$ and standard deviation $1$. Here we have considered the case of two trading intervals ($K=2$). The result for the optimal trading strategy derived through equation (\ref{ACutility}), $U^{{}_{\rm AC}}$, coincides the one obtained via the same equation (at this value of $\lambda=0.7$) considering normally-distributed returns (which is $(n_1^{\rm opt},n_2^{\rm opt}) = (0.85,0.15)$). Therefore AC gives the same optimal execution strategy in both cases (Gaussian and t-Student returns), and in this sense we say that AC is a ``mean-variance" approach to TCA. In contrast, the UF (\ref{DMUF}) gives, in this scenario, the optimal trading strategy $(n_1^{\rm opt},n_2^{\rm opt}) = (0.81,0.19)$. This is to be compared with the result $(n_1^{\rm opt},n_2^{\rm opt}) = (0.84,0.16)$ (c.f. table \ref{mapDMUDsols}) that is obtained using normally-distributed returns. The utility function (\ref{DMUF}) therefore ``recognises" the nature of TC, and it is in this sense that it is said to be a ``beyond mean-variance" approach to TCA.

It is instructive to analyse the optimal trading strategies obtained for Gaussian and non-Gaussian returns. While the result derived through the UF (\ref{ACutility}) is the same in both cases, equation (\ref{DMUF}) gives (for $\lambda=0.7$ and $\tilde{c} = -1$) the result $\vec{n}^{\rm opt} = (0.84,0.16)$ for Gaussian (arithmetic) returns, and $\vec{n}^{\rm opt} = (0.81,0.19)$ for t-Student (arithmetic) returns. Therefore in the second case (which, as we have said, is a better representation of reality) we ought to execute more patiently\footnote{Notice that the difference between both approaches corresponds, in this case, to imbalance of around $3\%$ of the shares to be traded more/less in each of the two time intervals.}. Is this intuitive? After all we may think that in a case in which we have more rare events (heavier tails), we should have a larger timing risk (under the picture of the UF (\ref{DMUF})) and thus, to minimise it, execute more aggressively than in the Gaussian case, not more patiently. This is indeed the situation when $\tilde{c}$ is sufficiently small, namely smaller than the threshold value $\sim {\rm mean}[c] - 1.89\times{\rm standard \; deviation}[c]$, which is the smallest point at which the Gaussian and t-Student cumulative distribution functions intersect. As $\tilde{c}$ is increased above this value, the value of the integral $\int_{-\infty}^{\tilde{c}} f[c]dc$ is actually greater for Gaussian TC than for t-Student TC, thus giving rise to the fact that in this case we ought to execute more patiently. Is is important to remember that $\tilde{c}$ allows us to modify the significance of timing risk, from extreme negative costs to less extremely negative (or even positive) TC. This conceptual variation, provided by the additional degree of freedom that $\tilde{c}$ in the UF (\ref{DMUF}) introduces, is what gave rise to the ``inversion effect" mentioned above (from executing more shares in earlier trading intervals than in later trading intervals to otherwise). In the future it would be interesting to investigate the role that autocorrelated returns (i.e. non-independently distributed) plays in the optimal trading strategy. We speculate that --for positive autocorrelation-- market impact gets amplified, therefore suggesting that in this case the order should be executed more patiently than for independent returns (and conversely for negative autocorrelation); but this is a topic for future work.

\newpage

\section{Alternative impact models}

We end this chapter by considering another area where the utility function (\ref{DMUF}) shows distinctively features. In section \ref{ProbdistSec} we have shown how --not only the nature of financial returns-- but also the conjectured impact model fundamentally affects the shape of the TC distribution. However, just as we can estimate the distributional properties of returns for a particular market realisation, market impact, as we have previously mentioned, cannot be measured: For a given security and point in time (unique market conditions), either we execute the order (in which case our own supply/demand affects the security's price, and this is what we observe), or we do not execute the order\footnote{More precisely, not only execution --but also sending any signal to the market of our willingness/desire to trade-- will affect the security's price as the market absorbs this information.} (in which case the security's price is not affected by our supply/demand, only by that of other market participants).
Since we cannot observe both instances of reality at the same time (either we execute the order or we do not), the difference between the security's price in the presence and in the absence of our execution cannot be measured. The best we can do in order to optimise TC is to estimate/predict MI based on a model. As such, having a realistic representation of reality (impact model) is of extreme importance --since as we will see-- the optimal trading strategy highly depends on the underlying assumptions of how execution affects a security's market price.

In this section we consider the price dynamics given by (\ref{DMimpactmodel}), as compared to the price evolution (\ref{pkAC}) considered so far. For the impact function ${\cal I}(n_k)$, we will contemplate three different possibilities:
\beq
{\rm LINEXP:} \quad {\cal I}(n_k) &=& -\xi \gamma n_k e^{-\rho \; k \Delta t}  \label{LINEXPeq} \\
{\rm LINPOW:} \quad {\cal I}(n_k) &=& -\xi \gamma n_k (k \Delta t)^{-\rho} \label{LINPOWeq} \\
{\rm SQRT:} \quad {\cal I}(n_k) &=& -\xi\gamma\sqrt{n_k} \label{SQRTeq}
\eeq
Where `LINEXP', `LINPOW', and `SQRT' stand for ``linear-exponential" (linear in the number of executed shares and exponential time decay), ``linear-power-law" (linear in the number of executed shares and power-law time decay), and ``square-root" (square root in the number of executed shares), respectively. These are some of the most popular functional forms of MI that appear in the literature within the propagator approach \cite{Gatheral09,Bouchaud18}. 
One important difference between the price evolution (\ref{DMimpactmodel}) and that of (\ref{pkAC}) is that, in the case that we now consider, both the impact law and the stochastic component are linked to {\it geometric} returns (as compared to arithmetic returns in the dynamics previously considered). This is relevant because: i) Geometric returns are generally better characterised by universal distributional properties (i.e. when we say that the returns follow a Gaussian or t-Student distribution). ii) Arithmetic returns are only a good approximation to geometric returns for small time increments. iii) With {\it geometric} returns TC are typically non-Gaussian, regardless on whether returns are normally distributed or not (since a product of stochastic variables --as involved in equation (\ref{TCIMDM})-- Gaussian or otherwise, is generally non-Gaussian). Our utility function (\ref{DMUF}) becomes therefore particularly relevant in this case: as we have seen, timing risk modelled as $\int_{-\infty}^{\tilde{c}} f[c]dc$ captures non-Gaussian features such as heavy tails, and in this case the optimal execution strategy depends on the nature of TC beyond mean-variance. This will be illustrated further now, by showing the difference between the result obtained via the AC utility function (\ref{ACutility}) (evaluated through numerical simulations only, as in this case no analytical expression for $\mathbb{E}[c]$ and $\mathbb{V}[c]$ is available) and via the UF (\ref{DMUF}).

\begin{table}[t]
\begin{center}
\resizebox{\columnwidth}{!}{
\begin{tabular}{|c|c|c||c|c|}
\cmidrule[1pt]{2-5}                  
\multicolumn{1}{c}{}&\multicolumn{2}{|c||}{{\bf Gaussian Rets.}}&\multicolumn{2}{|c|}{{\bf t-Student Rets.}} \\ 
\midrule
{\bf Impact model} & $\bm{\vec{n}^{\rm opt}}$ {\bf AC numeric} & $\bm{\vec{n}^{\rm opt}}$ \bf{DM} & $\bm{\vec{n}^{\rm opt}}$ {\bf AC numeric} & $\bm{\vec{n}^{\rm opt}}$ \bf{DM} \\\midrule
LINEXP & $(0.56,0.44)$ & $(0.29,0.71)$ & $(0.56,0.44)$ & $(0.28,0.72)$ \\
LINPOW & $(0.43,0.57)$ & $(0.33,0.67)$ & $(0.42,0.58)$ & $(0.32,0.68)$ \\
SQRT & $(0.43,0.57)$ & $(0.41,0.59)$ & $(0.43,0.57)$ & $(0.41,0.59)$ \\
\midrule
\end{tabular}
}
\caption[Optimal trading strategies for 2 trading intervals and different impact models.]{Solutions, expressed in the form $(n_1,n_2)$, for the optimal execution strategy obtained with various impact models, in the case in which the trading horizon is subdivided into of 2 time intervals ($K=2$). All results have been obtained by using the price evolution given in (\ref{DMimpactmodel}), with noise term $\zeta = \sigma \sqrt{\Delta t} \; \chi$, being $\chi\sim{\cal N}(0,1)$ --normally-distributed variable with mean $0$ and standard deviation $1$-- (`Gaussian Rets.') or $\chi\sim t_\nu(0,\sqrt{(\nu-2)/\nu}) \sqrt{\sigma})$ --t-Student-distributed variable with mean $0$, standard deviation $1$, and $\nu=5$ degrees of freedom-- (`t-Student Rets.').
LINEXP refers to `linear-exponential' (the solution obtained by using the impact model (\ref{LINEXPeq}) in the price evolution). 
LINPOW refers to `linear-power-law' (the solution obtained by using the impact model (\ref{LINPOWeq}) in the price evolution). 
SQRT refers to `square-root' (the solution obtained by using the impact model (\ref{SQRTeq}) in the price evolution). 
`AC numeric' denotes the solution (numerically) obtained by using the AC utility function (\ref{ACutility}). DM denotes the solution obtained by using the UF (\ref{DMUF}). The Monte Carlo sample contains $100$ quasi-random strategies in each case. Parameter values are $\gamma=1$, $\eta=1$, $\varepsilon=1$, $\sigma=1$, $\Delta t=1$, $\xi=1$, $p_0=1$, $\lambda = 0.3$, $\tilde{c} = -1$, $\rho=1/2$. The number of shares $n_k$, executed in time interval $k$, is expressed as a ratio to the order size, i.e. we take $N=1$. In these simulations, we have used $15000$ price paths to calculate each of $100$ trading strategies, which we fit to a cubic polynomial in order to obtain the optimal execution strategy.}
\label{TabOptSolsImpact2D}
\end{center}
\end{table}

\begin{figure}[h!]
\begin{center}
\includegraphics[width=0.9\linewidth]{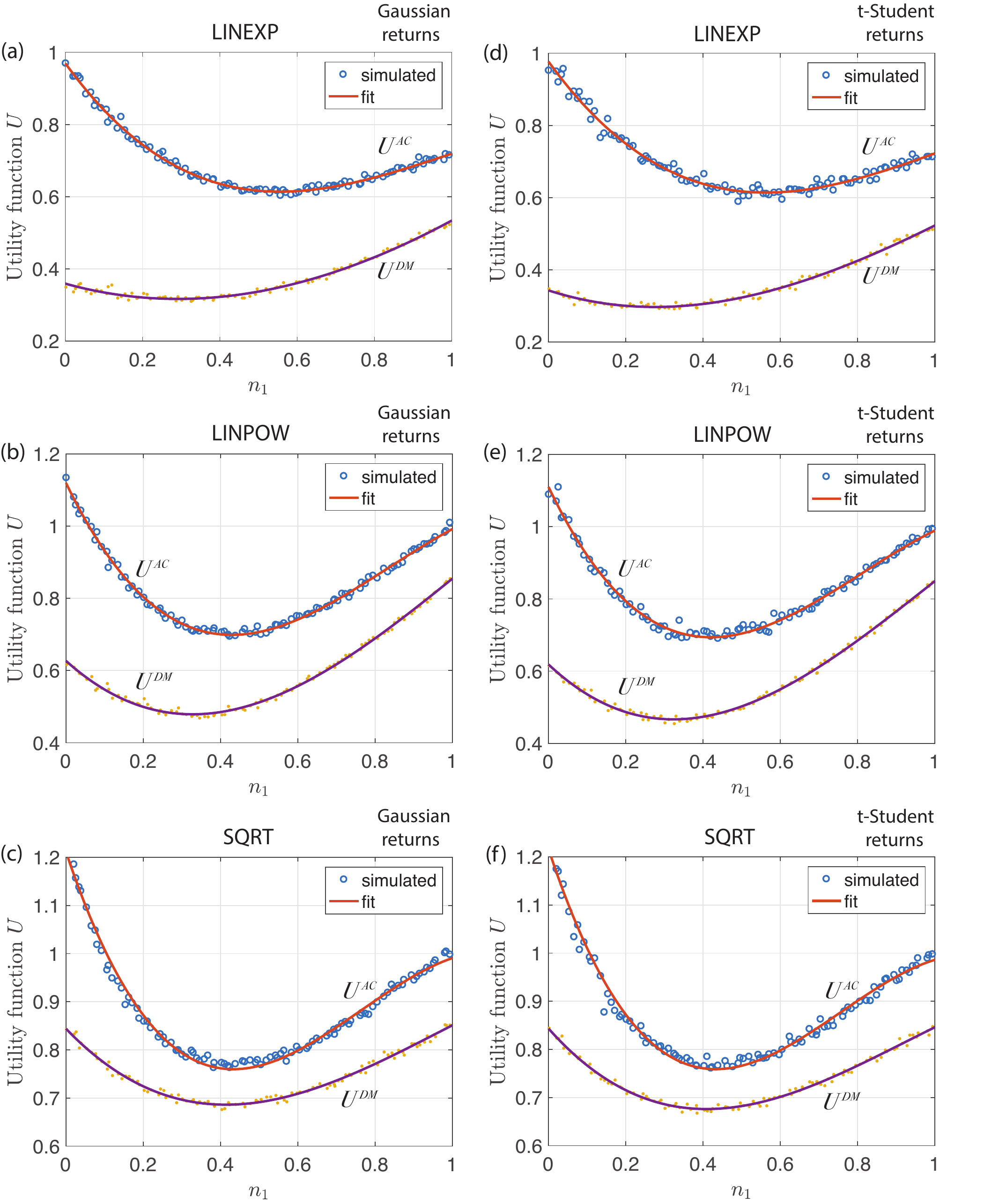}
\caption[Utility function for 2 trading intervals and different impact models.]{Utility function in the case in which the trading horizon is subdivided into of 2 time intervals ($K=2$), comparing different impact models, and Gaussian vs. t-Student returns. Here we show the utility functions (\ref{ACutility}), $U^{{}_{\rm AC}}$, and (\ref{DMUF}), $U^{{}_{\rm DM}}$, evaluated through the impact models LINEXP, LINPOW, SQRT (c.f. equations (\ref{LINEXPeq}), (\ref{LINPOWeq})), (\ref{SQRTeq})) and the price evolution (\ref{DMimpactmodel}). The minima of the UF (optimal trading strategies) correspond to the values shown in table \ref{TabOptSolsImpact2D}. Both utility functions have been numerically evaluated, using a Monte Carlo sample with $15000$ price paths to estimate each of the $100$ quasi-random strategies represented by the dots. The solid lines are a fit of a third-order polynomial to these points. Panels (a), (b) and (c) correspond to Gaussian (geometric) returns, while panels (d), (e) and (f) correspond t-Student (geometric) returns.
Parameter values are $\gamma=1$, $\eta=1$, $\varepsilon=1$, $\sigma=1$, $\Delta t=1$, $\xi=1$, $p_0=1$, $\lambda = 0.3$, $\tilde{c}=-1$, $\rho=1/2$. The number of shares $n_k$, executed in time interval $k$ is expressed as a ratio to the order size, i.e. we take $N=1$.}
\label{IMs_2D_ordpoly3}
\end{center}
\end{figure}

In table \ref{TabOptSolsImpact2D} we show the optimal trading strategies obtained with equations (\ref{ACutility}) and (\ref{DMUF}), in the case of 2 trading intervals ($K=2$), for each of the impact functions (\ref{LINEXPeq}), (\ref{LINPOWeq}), (\ref{SQRTeq}), and for both --Gaussian and t-Student-- returns. The new price evolution seems to have a significant impact on the optimal trading strategies. Most scenarios show an ``inversion effect'' with respect to the solutions derived through the time evolution (\ref{pkAC}). In the case of UF (\ref{DMUF}), TC are such that here the value of $\tilde{c}=-1$ incorporates ``not-so extreme" events in the timing-risk term. In the case of the UF (\ref{ACutility}) the effect is new. However, as we observe from the product over consequent time intervals in equation (\ref{TCIMDM}), there is now a major penalty for execution during earlier intervals, which dominates dominates over the noise term in transaction costs, contributing to the ``inversion effect" in all cases. This is true except for LINEXP evaluated with the UF (\ref{ACutility}), where `AC numeric' and `DM' in table \ref{TabOptSolsImpact2D} show $n_1>n_2$ and $n_2>n_1$, respectively. 
Again, the UF (\ref{DMUF}) tends to provide a solution in which more shares are executed in the latest trading intervals, a situation that is associated to the timing-risk term $\sim \int_{-\infty}^{\tilde{c}} f[c] dc$ incorporating both extreme and non-extreme TC.
Most interestingly, `AC numeric' provides the same optimal trading strategy for LINPOW and SQRT in this case, and `DM' executes most aggressively for SQRT.
We have also evaluated the case in which --for the price dynamics given by (\ref{DMimpactmodel}) and impact models (\ref{LINEXPeq}), (\ref{LINPOWeq}), (\ref{SQRTeq})-- (geometric) returns follow t-Student distribution. The optimal trading strategies, in this situation, are almost identical to the case of Gaussian returns. The impact models and price evolution considered here are therefore robust against certain distributional properties of financial returns (such as heavy tails). This characteristic holds at the level of the full utility function (not only for the optimal trading strategy), as it is shown in figure \ref{IMs_2D_ordpoly3}. For all impact models, the utility functions corresponding to Gaussian and t-Student returns are here almost identical. Furthermore, we see that now the quadratic approximation --previously used for the fitted UF-- is no longer appropriate. Instead, we have now used a {\it cubic} polynomial, which, as figure \ref{IMs_2D_ordpoly3} shows, provides a good fit to the simulated strategies in this case.
Although here the appropriate degree of the fitting polynomial has been assessed visually, and by comparison of the solutions obtained via Monte Carlo and a quadratic fit together with a gradient descent method, determining the pertinent functional form of the utility function should be done, in general, more rigorously. This may enter the territory of model selection and the well known problem of finding the optimal bias-variance tradeoff\footnote{That is, we need to find a good balance between goodness-of-fit and and the number of model parameters.} \cite{Hastie01}, a methodology that is beyond the scope of this text, but that might be worth investigating in the presence of more complex impact models or for a higher number of trading intervals.

~

~

~

~

~

~

~

~

~

~

~

~

~

~

~

~

~

~

~

~

~

~

~

~


\chapter{Conclusions} 
\label{Chapter4}
\lhead{Chapter 4. \emph{Conclusions}} 

\begin{flushright}

\textit{``You must unlearn what you have learned... Do. Or do not. There is no try."}

Yoda, Jedi Master

\end{flushright}

In this Thesis we have investigated transaction costs in securities trading. Specifically, we have developed a framework that enables us to characterise optimal execution of market orders in terms of two essential degrees of freedom: market impact and timing risk. Here --in contrast to previously existing theories that mathematically model these variables-- we have introduced an alternative representation of timing risk, namely it has been modelled as the TC incurred by our execution below a particular cost threshold. This characterisation of timing risk is embedded into a novel utility function, whose minimisation provides the optimal execution strategy. While the most common approaches to optimal execution rely on mean and variance of TC, only, our formalism captures features of the TC distribution beyond these first moments. This allows us to devise distinct optimal execution strategies for market situations whose associated TC are only differentiated by higher-order moments (such as the asymmetry of the distribution or the thickness of the tails of the distribution). More precisely, using our UF --encapsulated in equation (\ref{DMUF})-- we have shown that it is possible to have optimal execution strategies that trade most of the order volume in the latest time intervals of the trading horizon. 

~

Along the way towards establishing a formalism for optimal execution, we have encountered various difficulties. Namely, resolving the optimal trading strategy as the minimum of a utility function relies, in general, on a stochastic simulation of market prices. This inherently leads to non-deterministic, noisy utility functions, whose global minima are non-trivially determined. We have, nonetheless, developed a methodology that is sufficiently robust against statistical fluctuations, and by which a stable solution for the optimal execution strategy is reached at attainable levels of convergence of the TC sample to the underlying probability distribution. This framework has been used to resolve the optimal trading path when the time horizon is subdivided into a few intervals. By doing so, we have analysed in detail how the two control parameters of our utility function (``risk aversion" and ``cost threshold") determine the different ``phases" of the optimal trading strategy. With these foundations, we have addressed fundamental questions related to transaction costs, such as: i)~The relationship between the distributional properties of market returns and those of TC. ii)~The relationship between the functional form of price/impact models and the shape of the TC distribution. iii)~The relationship between market impact and timing risk: here we have shown that when timing risk is characterised as the (negative) tail of the TC distribution, market impact and timing risk follow an inverse {\it linear relationship}; something that is in stark contrast with the relationship that market impact and timing risk follow when the latter is modelled as TC variance. 

Our inquiry on execution trading has been complemented with the development and analysis of a price-impact model, which lead to the equation (\ref{TCIMDM}). As we have shown, this has distinctive characteristics. Among them is the fact that execution impatience is penalised, i.e. it allows solutions to optimal execution strategies that trade the majority of the order shares in the latest time intervals. Furthermore, it is robust with respect to different TC distributions, namely the optimal trading strategies and utility functions obtained for Gaussian and t-Student returns are, in this case, almost indistinguishable from each other. We have analysed in detail three different functional forms of the impact function, showing in particular how a sub-linear dependence on the executed volume tends to give ``more impatient" optimal trading strategies. Finally, we have examined important applications of our theory, such as the case when market returns include features beyond mean-variance. In particular, by analysing the case in which returns follow a t-Student distribution, we have identified two regimes (controlled by the value of the cost threshold) in which heavy tails contribute to more or less aggressive execution strategies.

~

Our work opens new questions and future prospects. Among them we can cite:
\begin{enumerate}[i)]
\item The scaling of the formalism to higher dimensionality (i.e. larger number of trading intervals).
\item The treatment of the trading horizon as an optimisation variable, and its relationship to the risk-aversion parameter.
\item Extending the framework to include opportunity cost, i.e. the case in which the order might be incompletely executed by the end of the trading horizon.
\item The application of our utility function to other impact models.
\item Generalising the formalism to treat market-on-close orders and limit orders.
\item Analysing the effect of other utility functions of the class introduced here, such as (\ref{UFclass1}) and (\ref{UFclass2}), and variations of them in order to recover the ``U-shape" common in daily executed volumes.
\item Formalise further the process by which the appropriate fitted utility function is determined, namely by employing model selection and goodness-of-fit tests as the dimensionality is increased.
\end{enumerate}

Finally, from a broader perspective, it would be interesting to develop a formalism that unifies optimal execution with portfolio optimisation. In particular, when the timescales of portfolio rebalancing are on the same order of those involved in order execution, a common framework that dynamically determines the weights of a portfolio --while the optimal trading trading paths for these securities are simultaneously established in order to maximise the PnL-- is a thrilling area of research where the work developed here could find especial relevance.

~

~

~

~

~

~

~

~

~

~

~

~

~

~

~

~

~

~

~

~

~

~

~

~

~

~

~

~

~

~

~

~

~

~

~

~

~

~

~

~


\addtocontents{toc}{\vspace{1em}} 

\appendix 


\chapter{Fundamental benchmarks and trading algorithms}
\label{AppBenchmarks}
\lhead{Appendix A. \emph{Fundamental benchmarks and trading algorithms}}

\subsection*{TWAP (Time Weighted Average Price)}
 
This strategy trades `uniformly' over the time horizon $T$, i.e.
\beq
n_k^{\rm TWAP} = \frac{N}{K}
\eeq
Trading at times $t_k = \frac{T}{K} k$, where $k = 1,\ldots,K$. Here $N$ is the total number of shares to be traded and $K$ the number of time sub-intervals over the time $T$. The benchmark price is given by
\beq
p^{\rm TWAP} = \frac{1}{K}\sum_{k=1}^N p_k
\eeq
Where $p_k$ is the average transaction price in time bin $k$.
 
\subsection*{VWAP (Volume Weighted Average Price)}
 
This strategy trades proportionally to the market volume over the time horizon $T$. If $\nu_k$ is the traded market volume\footnote{Notice that for VWAP as an algorithm, $\nu_k$ will be a {\it predicted} volume from historical observations, while for VWAP as a benchmark, $\nu_k$ corresponds to the {\it realised} market volume.} over time bin $k$, we have
\beq
n_k^{\rm VWAP} = \frac{\nu_k}{V} N
\eeq
Here $V$ is the market volume traded over the time period $T$ and $N$ the total number of shares to be traded. The benchmark price is given by
\beq\label{TCAVWAP}
p^{\rm VWAP} = \frac{1}{V}\sum_{k=1}^N \nu_k p_k
\eeq
Where $p_k$ is the average transaction price in time bin $k$.
 
\subsection*{POV (Percentage of Volume) \& \\ PWP (Participation Weighted Price)}
 
This strategy trades according to a percentage of the traded market volume in each time bin. A POV algorithm is accompanied by a `participation rate' parameter $\eta$ (with $0\leq \eta \leq 1$), and the number of shares traded in time bin $k$ is given by
\beq
n_k^{\rm POV} = \eta \nu_k,
\eeq
where $\nu_k$ is the traded market volume over time bin $k$. The associated benchmark is called PWP, whose price is given by
\beq\label{TCAPWP}
p^{\rm PWP} = \eta \sum_{k=1}^N \nu_k p_k
\eeq
 
\subsection*{MO (Market Open)}
 
MO only acts as a benchmark (so $n_k^{\rm bmk}=n_k^{\rm exe}$), given a trading strategy $\{n_k^{\rm exe}\}$. The benchmark price is simply the price of the security at the market opening time\footnote{For discrete times this is taken to be the price of the security at the beginning of the corresponding time bin.}
\beq
p^{\rm IS} \equiv p_O
\eeq
 
\subsection*{MC (Market Close)}
 
MC only acts as a benchmark (so $n_k^{\rm bmk}=n_k^{\rm exe}$), given a trading strategy $\{n_k^{\rm exe}\}$. The benchmark price is simply the price of the security at the market closing time\footnote{For discrete times this is taken to be the price of the security at the end of the corresponding time bin.}
\beq
p^{\rm IS} \equiv p_C
\eeq
 
\subsection*{IS (Implementation Shortfall)}
 
Defined as a benchmark, IS is simply the price of the security at the order start time\footnote{For discrete times this is taken to be the price of the security at the start or end (a matter of criterium) of the corresponding time bin.}
\beq
p^{\rm IS} \equiv p_0
\eeq
So, taking into account that $\sum_{k=1}^K n_k^{\rm bmk} = N$, the transaction cost associated to this benchmark can be calculated as
\beq \label{ISdefEq}
c_{\rm IS} \equiv \frac{\xi}{N} \sum_{k=1}^K \left( n_k^{\rm exe} p_k^{\rm exe} - p_0 \right),
\eeq
where $\xi =-1$ for a buy order and $\xi =1$ for a shell order. Defined as an algorithm, IS tries to minimise a utility function of transaction costs, as described in the main text. 

It is worth noticing that an alternative conception of IS as a benchmark is as follows: compute the predicted cost, given equation (\ref{ISdefEq}), predicted prices (determined by historical observations), and the corresponding optimal trading strategy. This value, which we will call {\it estimated-IS}, can be compared to the observed IS calculated from equation (\ref{ISdefEq}), given the executed strategy and the corresponding execution prices.
 
~

~

~

~

~

~

~

~

~

~

~

~

~

~

~

~

~

~

~

\addtocontents{toc}{\vspace{1em}} 

\clearpage  

\addtocontents{toc}{\vspace{0em}}  
\backmatter
\label{Bibliography}
\lhead{\emph{Bibliography}}  
\bibliographystyle{alpha}  
\bibliography{Bibliography}  

\end{document}